\documentclass[%
superscriptaddress,
 amsmath,amssymb,
prb,twocolumn,
]{revtex4-2}
\usepackage{siunitx}
\usepackage{graphicx}% Include figure files
\usepackage{dcolumn}% Align table columns on decimal point
\usepackage{bm}% bold math
\usepackage{hyperref}% add hypertext capabilities
\usepackage[%showframe,%Uncomment any one of the following lines to test 
letterpaper,margin=1in]{geometry}

\usepackage{xcolor}
\usepackage{multirow}
\usepackage[inline]{enumitem}
\setlist[itemize]{leftmargin=*}
\setlist[enumerate]{leftmargin=*}
\setlist{noitemsep}
\DeclareSIUnit\angstrom{\text {Å}}
\DeclareSIUnit\torr{Torr}

\begin{document}

% \preprint{APS/123-QED}
%TC:ignore
\title{Hydrogen is not necessary for superconductivity in topotactically reduced nickelates}% Force line breaks with \\

\author{Purnima P. Balakrishnan}%
 \email{purnima.balakrishnan@nist.gov}
\affiliation{%
NIST Center for Neutron Research, National Institute of Standards and Technology, Gaithersburg, MD 20899, USA
}%

 \author{Dan Ferenc Segedin}
\affiliation{%
 Department of Physics, Harvard University, Cambridge, MA 02138, USA
}%

\author{Lin Er Chow}
\affiliation{
Department of Physics, Faculty of Science, National University of Singapore, Singapore 117551, Singapore
}

\author{P. Quarterman}
\affiliation{%
NIST Center for Neutron Research, National Institute of Standards and Technology, Gaithersburg, MD 20899, USA
}%

\author{Shin Muramoto}
\affiliation{
 Material Measurement Laboratory, National Institute of Standards and Technology, Gaithersburg, MD 20899, USA
}%

\author{Mythili Surendran}
\affiliation{%
 Mork Family Department of Chemical Engineering and Materials Science, University of Southern California, Los Angeles, CA 90089, USA}
\affiliation{
Core Center for Excellence in Nano Imaging, University of Southern California, Los Angeles, CA 90089, USA
}%

\author{Ranjan K. Patel}
\affiliation{%
 Department of Physics, Indian Institute of Science, Bengaluru, 560012, India
}%

\author{Harrison LaBollita}
\affiliation{Department of Physics, Arizona State University, Tempe, AZ 85287, USA}

\author{Grace A. Pan}
\affiliation{%
 Department of Physics, Harvard University, Cambridge, MA 02138, USA
}%

\author{Qi Song}
\affiliation{%
 Department of Physics, Harvard University, Cambridge, MA 02138, USA
}%

\author{Yang Zhang}
\affiliation{The Rowland Institute at Harvard, Harvard University, Cambridge, MA 02138, USA}

\author{Ismail El Baggari}
\affiliation{The Rowland Institute at Harvard, Harvard University, Cambridge, MA 02138, USA}

\author{Koushik Jagadish }
\affiliation{Mork Family Department of Chemical Engineering and Materials Science, University of Southern California, Los Angeles, CA 90089, USA}

\author{Yu-Tsun Shao}
\affiliation{Mork Family Department of Chemical Engineering and Materials Science, University of Southern California, Los Angeles, CA 90089, USA}

\author{Berit H. Goodge}
\affiliation{School of Applied and Engineering Physics, Cornell University, Ithaca, New York 14853,USA}
\affiliation{Kavli Institute at Cornell for Nanoscale Science, Ithaca, New York 14853, USA}
\author{Lena F. Kourkoutis}
\affiliation{School of Applied and Engineering Physics, Cornell University, Ithaca, New York 14853,USA}
\affiliation{Kavli Institute at Cornell for Nanoscale Science, Ithaca, New York 14853, USA}

\author{Srimanta Middey}
\affiliation{%
 Department of Physics, Indian Institute of Science, Bengaluru, 560012, India
}%

\author{Antia S. Botana}
\affiliation{Department of Physics, Arizona State University, Tempe, AZ 85287, USA}

\author{Jayakanth Ravichandran}
\affiliation{%
Mork Family Department of Chemical Engineering and Materials Science, University of Southern California, Los Angeles, CA 90089, USA}
\affiliation{
Core Center for Excellence in Nano Imaging, University of Southern California, Los Angeles, CA 90089, USA
}%
\affiliation{Ming Hsieh Department of Electrical and Computer Engineering, University of Southern California, Los Angeles, CA 90089, USA
}%

\author{A. Ariando}
\affiliation{
Department of Physics, Faculty of Science, National University of Singapore, Singapore 117551, Singapore
}

\author{Julia A. Mundy}
\email{mundy@fas.harvard.edu}
\affiliation{%
 Department of Physics, Harvard University, Cambridge, MA 02138, USA
}%

\author{Alexander J. Grutter}
 \email{alexander.grutter@nist.gov}
\affiliation{%
 NIST Center for Neutron Research, National Institute of Standards and Technology, Gaithersburg, MD 20899, USA
}%

\date{\today}% It is always \today, today,

\begin{abstract}

A key open question in the study of layered superconducting nickelate films is the role that hydrogen incorporation into the lattice plays in the appearance of the superconducting state. Due to the challenges of stabilizing highly crystalline square planar nickelate films, films are prepared by the deposition of a more stable parent compound which is then transformed into the target phase \textit{via} a topotactic reaction with a strongly reducing agent such as CaH$_2$. Recent studies, both experimental and theoretical, have introduced the possibility that the incorporation of hydrogen from the reducing agent into the nickelate lattice may be critical for the superconductivity. In this work, we use secondary ion mass spectrometry to examine superconducting La$_{1-x}$$X_{x}$NiO$_{2}$ / SrTiO$_{3}$ ($X$ = Ca and Sr) and Nd$_{6}$Ni$_{5}$O$_{12}$ / NdGaO$_{3}$ films, along with non-superconducting NdNiO$_2$ / SrTiO$_3$ and (Nd,Sr)NiO$_2$ / SrTiO$_3$. We find no evidence for extensive hydrogen incorporation across a broad range of samples, including both superconducting and non-superconducting films. Theoretical calculations indicate that hydrogen incorporation is broadly energetically unfavorable in these systems, supporting our conclusion that hydrogen incorporation is not generally required to achieve a superconducting state in layered square-planar nickelates.

\end{abstract}

\maketitle
%TC:endignore
%\tableofcontents

% \section{Introduction}

\begin{figure*}
    \centering
    \includegraphics[width=\textwidth]{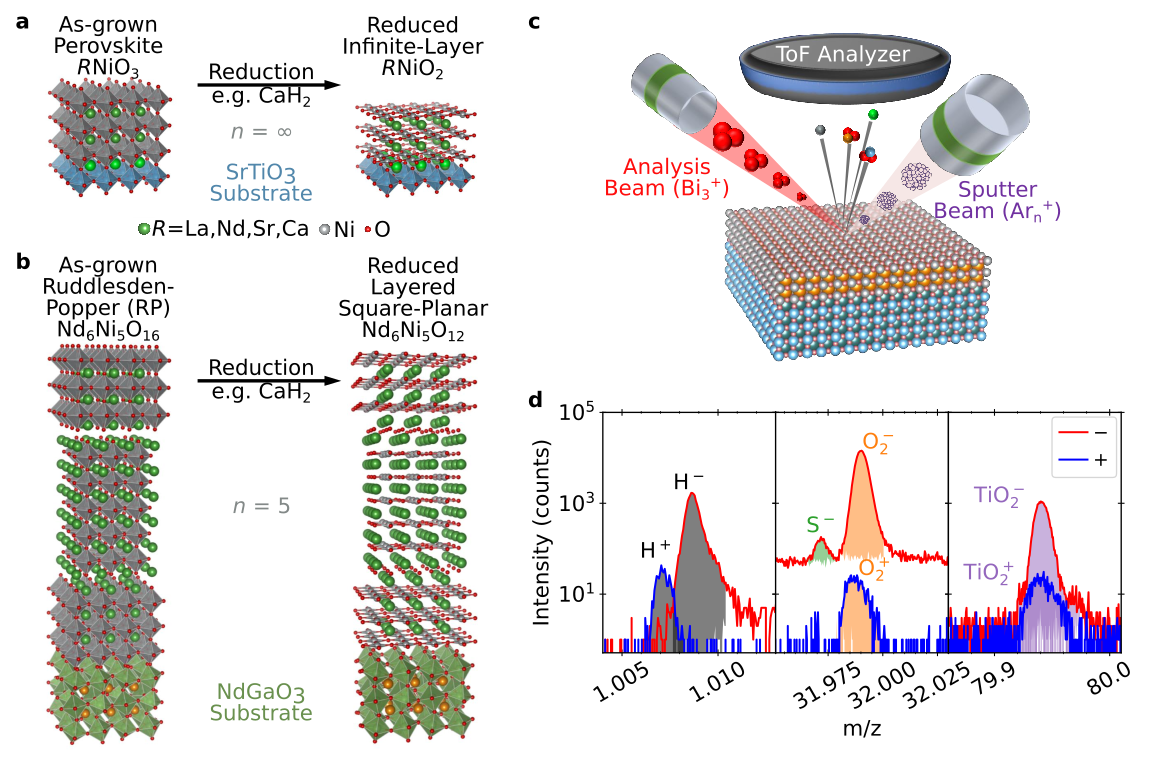}
    \caption{Schematic crystal structures for precursor phase and reduced a. $n$ = $\infty$ and b. $n$ = 5 layered square-planar nickelate compounds. c. Schematic of the ToF-SIMS measurement technique. d. SIMS spectra measured separately for positive and negative ions are analyzed by identifying peaks by mass to charge ratio ($m/z$) and extracting integrated area.
    }
    \label{fig:schematics}
\end{figure*}

Superconductivity in nickelates has been pursued ever since the discovery of the cuprates \cite{Anisimovtheory1999,LNOLMOSLtheory2008,LNOLAOtheory2009,Theory3D2010,SuperlatticeTheory2011}, but it was not until 2019 that it was demonstrated in thin films of the infinite-layer compound NdNiO$_2$ via hole doping with Sr \cite{LiDiscovery2019}. This discovery introduced a novel family of layered nickelate superconductors that has now been extended to include the Pr- and La- analogs of the infinite-layer compound as well as the five-layer material Nd$_6$Ni$_5$O$_{12}$ \cite{OsadaPSNO2020,osada2020PSNOphasediagram,WeiScavengingSC2023,pan2022n5RP}. While superconducting nickelates exhibit many interesting phenomena \cite{FowlieMagnetism2022,Lee2023,LaneTheory2023,lu2021magnetic}, they also represent a unique materials synthesis challenge \cite{ZHOU2022170,ferenc2023limits,lee2020aspects,parzyck2024synthesis,PhysRevMaterials.6.055003}. In general, layered square-planar nickelates cannot be synthesized directly, instead requiring a two-step fabrication method wherein an oxygen-rich precursor material is grown by traditional thin film deposition methods and then topotactically reduced, as illustrated in Fig. \ref{fig:schematics}a,b. Typically, reduction is performed \textit{via} a thermal anneal employing a chemical reducing agent and oxygen sink, such as H$_2$, NaH, or CaH$_2$ \cite{parzyck2024absence, WeiScavenging2023,Hayward1999, LiDiscovery2019}. One of the most pressing open questions is the degree to which the reduction process incorporates hydrogen into the nickelate film, and whether hydrogen is important in stabilizing superconductivity.

A notable recent study by X. Ding \textit{et al.} reported that hydrogen is critical for the emergence of superconductivity, requiring a stoichiometry around Nd$_{0.8}$Sr$_{0.2}$NiO$_2$H$_{0.25}$ \cite{DingHydrogen2023}. However, in this study, the precursor films are hydrogen-rich, while hydrogen and oxygen stoichiometry are highly correlated through reduction. Furthermore, recent results using alternative reduction processes with no hydrogen source have cast doubt on the role of incorporated hydrogen on the electronic state \cite{WeiScavenging2023,WeiScavengingSC2023}. Theoretical investigations yield widely varying results, with conclusions ranging from hydrogen incorporation promoting  \cite{SiHighConcentrations2023}, having no effect \cite{Cataldo2023}, or suppressing superconductivity altogether \cite{SiTopotactic2020}.

Given the differing conclusions in the literature, a comprehensive examination of the role of hydrogen incorporation in superconducting nickelates is urgently needed. To understand more broadly applicable trends rather than the specifics of one sample type or fabrication protocol, we used time-of-flight secondary ion mass spectrometry (ToF-SIMS) to study the relationship between hydrogen incorporation and superconductivity in a broad range of nickelate films grown and reduced by three different research groups. The films used in this study were grown \textit{via} either molecular beam epitaxy (MBE) or pulsed laser deposition (PLD), in a variety of geometries, and reduced with CaH$_2$ at different conditions. As the energetics of incorporating hydrogen may vary greatly depending on stoichiometry and structure \cite{SiTopotactic2020}, we compared multiple nickelate systems, including superconducting examples of La$_{1-x}$Ca$_x$NiO$_2$, La$_{1-x}$Sr$_x$NiO$_2$, and Nd$_6$Ni$_5$O$_{12}$. We also examined non-superconducting NdNiO$_2$ and Nd$_{1-x}$Sr$_x$NiO$_2$. We find no evidence that a large concentration of incorporated hydrogen is necessary to observe superconductivity. Instead, a wide range of films, superconducting and non-superconducting, exhibit H$^-$ intensities which are similar to the substrate background. Theoretical calculations support this picture, revealing that hydrogen incorporation is energetically unfavorable across all materials studied in this work.

As illustrated in Fig. \ref{fig:schematics}c, ToF-SIMS is a destructive technique in which an ion beam sputters through the film, and ejected molecular ions are analyzed using a mass spectrometer to provide a depth- and element-resolved picture of the ejected species, and thereby chemical composition. ToF-SIMS allows the isolation and identification of elemental H$^\pm$, and O$^\pm$ as well as larger ejected molecules such as OH$^\pm$, O$_2^\pm$, and TiO$_2^\pm$, as shown in Fig. \ref{fig:schematics}d. The change in molecular species intensity over time as the sample is sputtered results in a nanometer-resolved understanding of the chemical composition with depth.

However, the measured intensity depends significantly on the sputtering conditions, chemical environment, film composition, density, electronic state, and prevalence of structural defects. Absolute scaling of stoichiometry and depth therefore requires calibration standards with known stoichiometry and similar chemical environment to the films of interest. Since the defect levels and chemistry in superconducting nickelates evolves extensively during the reduction process, such standards are nearly impossible to obtain, and we instead adopt the convention of Ding \textit{et al.}, in which the hydrogen level observed in the SrTiO$_3$ or NdGaO$_3$ substrate is considered to be the ``background" level representing minimal hydrogen \cite{DingHydrogen2023}.

\section{\texorpdfstring{Superconducting L\MakeLowercase{a}$_{1-x}$(S\MakeLowercase{r},C\MakeLowercase{a})$_{x}$N\MakeLowercase{i}O$_2$}{Superconducting La1-x(Sr,Ca)xNiO2}}

\begin{figure*}
    \centering
    \includegraphics[width=\textwidth]{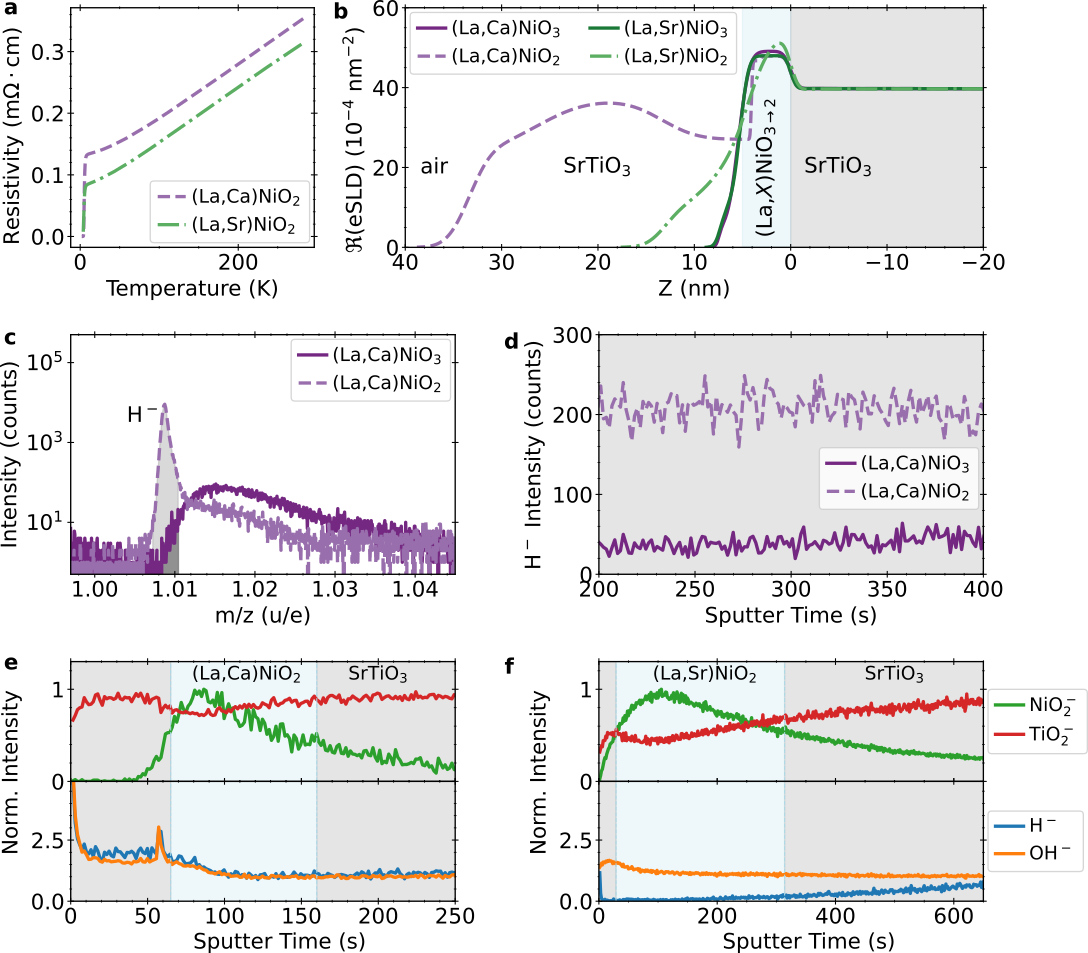}
    \caption{a. Resistivity vs. temperature for the superconducting La$_{0.78}$Ca$_{0.22}$NiO$_2$ and La$_{0.8}$Sr$_{0.2}$NiO$_2$ samples, showing a clear transition and large RRR. b. XRR depth profiles of the as-grown and superconducting films, specifically the real component of the scattering length density (SLD). c. Intensity vs. mass to charge ratio near the H$^-$ peak for an as-grown La$_{0.78}$Ca$_{0.22}$NiO$_3$ and superconducting La$_{0.78}$Ca$_{0.22}$NiO$_2$ film, integrated over the entire meaurement time. d. Raw intensity (counts) of the H$^-$ peak in the substrate region for the same as-grown La$_{0.78}$Ca$_{0.22}$NiO$_3$ and superconducting La$_{0.78}$Ca$_{0.22}$NiO$_2$. e. SIMS depth profile of superconducting La$_{0.78}$Ca$_{0.22}$NiO$_2$. f. SIMS depth profile of superconducting La$_{0.8}$Sr$_{0.2}$NiO$_2$.} 
    \label{fig:SCperovskite}
\end{figure*}

We first present results from two doped superconducting infinite-layer systems: La$_{0.78}$Ca$_{0.22}$NiO$_2$ and La$_{0.8}$Sr$_{0.2}$NiO$_2$ grown by pulsed laser deposition on SrTiO$_3$ substrates, as described in section \ref{sec:LXNOgrowth}. The quality of representative samples has been previously demonstrated through x-ray diffraction (XRD), cross-sectional scanning transmission electron microscopy (STEM), and electron energy loss spectroscopy (EELS) analysis \cite{chow2023pairing,chow2022paulilimit}. To ensure depth-wise uniformity, the film thickness was limited to below \qty{6}{\nano\meter}. Fig. \ref{fig:SCperovskite}a shows the superconducting transitions for these samples, which have a residual resistivity ratio $\approx$ 4.1, comparable to the highest reported values so far \cite{parzyck2024synthesis,sunadvmat}. After reduction, an amorphous SrTiO$_3$ cap is deposited to act as an oxidation barrier, with varying thickness due to the challenges associated with the room-temperature growth. Film and cap thicknesses were verified using x-ray reflectometry (XRR), shown in Fig. \ref{fig:SCperovskite}b, which reveals that the initial perovskite phase is uniform with the expected scattering length densities. After reduction, the sharp interfaces slight roughen, likely linked to the energetic deposition of the caps. While we focus on superconducting films, we also measured the as-grown perovskite and over-reduced (non-superconducting) films from the same growth, the details of which can be found in the supplemental information section \ref{sec:allSIMS}.

To understand the sensitivity of this experiment to hydrogen, we first compare the overall hydrogen content of as-grown La$_{0.78}$Ca$_{0.22}$NiO$_3$ and superconducting La$_{0.78}$Ca$_{0.22}$NiO$_2$ samples in Fig. \ref{fig:SCperovskite}c. Here, we show the H$^-$ peak for both samples integrated across all sputtering times. The as-grown sample contains negligible hydrogen within either the film or substrate, indicating an extremely clean growth and handling process. The superconducting sample, in contrast, exhibits a clearly-resolved H$^-$ peak much larger than the measurement background. The additional hydrogen introduced into the reduced superconducting sample is easily detectable. We next compare the integrated peak intensity, indicated by the shaded region, over sputter time (depth). For this same pair of samples, Fig. \ref{fig:SCperovskite}d shows the H$^-$ intensity in just the SrTiO$_3$ substrate. Interestingly, while the substrate intensity of the as-grown sample is less than 40 counts per frame, the SrTiO$_3$ substrate associated with the superconducting La$_{0.78}$Ca$_{0.22}$NiO$_2$ is an excellent match for the samples in Ding \textit{et al.} \cite{DingHydrogen2023}, with approximately 200 counts per frame of H$^-$. Thus we may be confident that the observed hydrogen levels in the substrates are above the instrumental detection limit and closely match previous observations.

Having firmly established that the hydrogen levels reported previously in superconducting films are readily detectable with the instrument used in this study, we show SIMS data from superconducting La$_{0.78}$Ca$_{0.22}$NiO$_2$ and La$_{0.8}$Sr$_{0.2}$NiO$_2$ in Fig. \ref{fig:SCperovskite}e,f. Here the NiO$_2$ and TiO$_2$ intensities are normalized to their maximum values while the H$^-$ and OH$^-$ intensities are normalized to the steady-state value within the substrate, while alternative normalizations and raw counts are shown in the supplemental information section \ref{sec:allSIMS}. The film and substrate positions are indicated by the peak and dip in NiO$_2^-$ and TiO$_2^-$ intensities, respectively. The trends in H$^-$ and OH$^-$ intensities quite clearly disagree with prior reports: the superconducting La$_{0.78}$Ca$_{0.22}$NiO$_2$ and La$_{0.8}$Sr$_{0.2}$NiO$_2$ films do not exhibit the large 1 to 3 order of magnitude increases in H$^-$ or OH$^-$ signal which would be expected for extensive, multiple-percent hydrogen incorporation.

Instead, apart from the quickly decaying signal associated with surface adsorbates, the H$^-$ and OH$^-$ signals within the La$_{0.78}$Ca$_{0.22}$NiO$_2$ film are invariant within a factor of two of the signal within the substrate. Interestingly, the La$_{0.78}$Ca$_{0.22}$NiO$_2$ sample, with a thicker amorphous SrTiO$_3$ cap (29 nm), exhibits higher H$^{-}$ intensity within the cap than within the nickelate film, concentrated near the interface. In the La$_{0.8}$Sr$_{0.2}$NiO$_2$ sample with a thinner SrTiO$_3$ cap (approximately 6 nm), H$^-$ is much lower in the nickelate film than either the SrTiO$_3$ substrate or the other superconducting film. We speculate that the SrTiO$_3$ cap may play a role in hydrogen capture or transport \cite{STOhydrogen}. Most importantly, the coexistence of different low hydrogen concentrations with superconductivity definitively demonstrates that hydrogen doping is not required for superconductivity in the infinite-layer nickelates.

\section{\texorpdfstring{Superconducting N\MakeLowercase{d}$_6$N\MakeLowercase{i}$_5$O$_{12}$}{Superconducting Nd6Ni5O12}}

\begin{figure}
    \centering
    \includegraphics[width=\columnwidth]{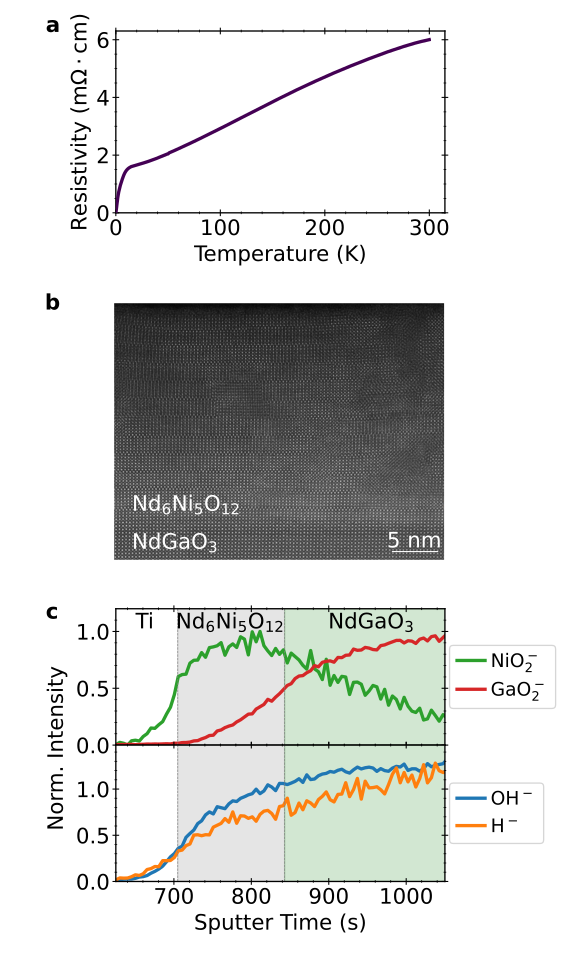}
    \caption{a. Temperature-dependent resistivity of reduced Nd$_{6}$Ni$_{5}$O$_{12}$ / NdGaO$_{3}$ showing a clear superconducting transition. Reproduced from Ref. \cite{pan2022n5RP}. b. STEM image of the reduced, superconducting Nd$_6$Ni$_5$O$_{12}$ film and NdGaO$_3$ substrate. c. SIMS depth profiles of superconducting Nd$_{6}$Ni$_{5}$O$_{12}$.
    }
    \label{fig:SCRP}
\end{figure}

To test whether our findings are applicable more broadly within the square-planar nickelate family, beyond the infinite-layer structure, we examine the $n=5$ member: superconducting Nd$_6$Ni$_5$O$_{12}$. This film consists of 23 nm Nd$_6$Ni$_5$O$_{12}$ synthesized on NdGaO$_{3}$ (110) (see section \ref{sec:NNOgrowth} for details), with 10 nm titanium followed by 100 nm platinum patterned on the film surface as electrodes. Fig. \ref{fig:SCRP}a shows the zero-field superconducting transition of this sample from reference \cite{pan2022n5RP}, with a residual resistivity ratio of 3.8. Further characterization of this sample can be found in Ref. \cite{pan2022n5RP}. Fig. \ref{fig:SCRP}(b) shows a representative STEM image of this sample, revealing the $n = 5$ square-planar structure. 

Fig. \ref{fig:SCRP}c plots the SIMS depth profile of this superconducting sample.  NiO$_2^-$ and GaO$_2^-$ peaks are normalized to their maximum values, and clearly identify the electrode, film, and substrate regions. As before, we obtain information regarding the hydrogen concentration by examining the relative intensity of the H$^-$ and OH$^-$ peaks in the film and substrate. An advantage of the relatively thick electrode is the removal of surface contaminant effects from the measurement. Both the H$^-$ and OH$^-$ intensities are low in the platinum and titanium, increase slowly in the Nd$_6$Ni$_5$O$_{12}$ film, and further increase deeper into the NdGaO$_3$ substrate. Similar to the superconducting La$_{0.8}$Sr$_{0.2}$NiO$_2$ sample, we find that the hydrogen content appears to be highest in the substrate, although again the nickelate film and substrate intensities are very similar. Once again there is no evidence of an order of magnitude increase in hydrogen intensity in the film.

\section{Non-superconducting Films}

With little evidence of extensive hydrogen incorporation in high-quality superconducting samples, the question remains whether some structures or processes are more susceptible to hydrogen. We speculate that films with increased defect densities, whether due to growth conditions or from long or overly aggressive reduction treatments, may incorporate additional hydrogen as a defect compensation mechanism. These films do not exhibit superconductivity, but do provide a mechanism for understanding the extent to which hydrogen can be incorporated during reduction and whether it might inhibit the fabrication of superconducting films. 

\begin{figure*}
    \centering
    \includegraphics[width=\textwidth]{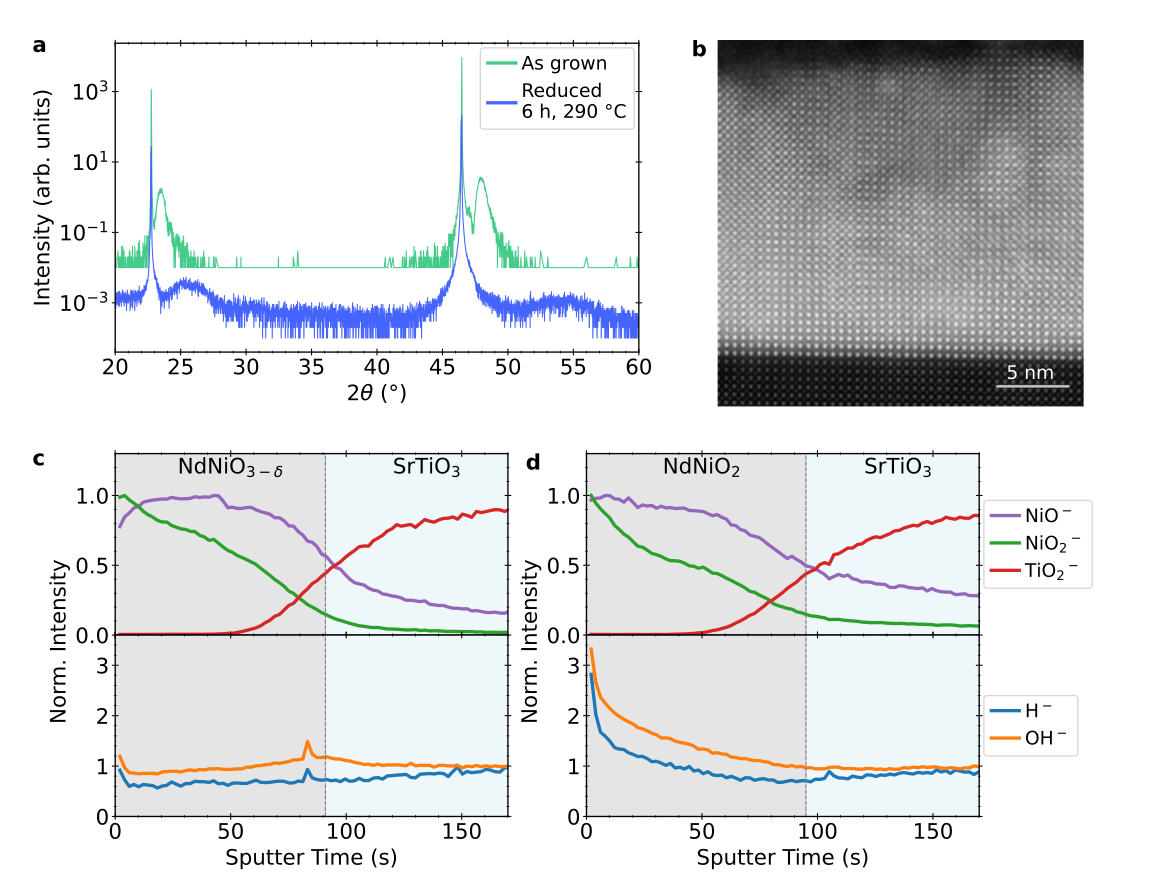}
    \caption{a. XRD of as-grown and NdNiO$_3$ and reduced NdNiO$_2$ films showing partial reduction to the infinite-layer phase. b. STEM of an equivalent sister NdNiO$_2$ sample revealing extended defects concentrated near the surface.  c. SIMS depth profile of the as-grown film NdNiO$_3$ film. d. SIMS depth profile of the NdNiO$_2$ film, reduced for an extended time of \qty{6}{\hour} at \qty{290}{\degreeCelsius}. SIMS shows increased hydrogen concentration at the surface even without full oxygen removal.}
    \label{fig:NNO-overreduced}
\end{figure*}

We first consider \qty{17}{\nano\meter} NdNiO$_3$ / SrTiO$_{3}$ (001) films grown by MBE and subjected to an incomplete reduction, at a lower temperature but for longer times compared to the optimized treatment for achieving high-quality NdNiO$_2$. XRD scans shown in Fig. \ref{fig:NNO-overreduced}a indicate reduction towards the infinite-layer NdNiO$_2$ phase, but with a modest decrease in crystallinity. Electron microscopy measurements on an equivalent sister sample, shown in Fig. \ref{fig:NNO-overreduced}b, reveals the presence of defects and phase boundaries, as expected.

The film was cut in half before reduction, and both as-grown and reduced samples were measured using ToF-SIMS, yielding the intensity depth profiles in Figures \ref{fig:NNO-overreduced}c,d. As before, the NiO, NiO$_2$, and TiO$_2$ peaks are normalized to their maximum value while the H$^-$ and OH$^-$ are normalized to the steady-state values in the substrate. The H$^-$ and OH$^-$ signals are slightly higher in the SrTiO$_3$ substrate than in the as-grown NdNiO$_3$ film. Upon reduction, H$^-$ and OH$^-$ increase at the surface of the films, and the lineshape of this increase only partially matches that of various peaks including C$_2^-$ and Ca$^+$ (see supplemental section \ref{sec:adsorbates}), indicating that they do not solely originate from surface adsorbates introduced during the reduction process. %, likely adventitious hydrocarbons and CaH$_2$ residue.
Near the substrate interface, which has previously been shown to be the highest-quality region of the film \cite{lee2020aspects,chow2023pairing}, the H$^-$ intensity remains lower than in the SrTiO$_3$ substrate. Thus, while some insignificant hydrogen content may be introduced during the reduction process, it seems to be limited near the surface of these uncapped films.

\begin{figure*}
    \centering
    \includegraphics[width=\textwidth]{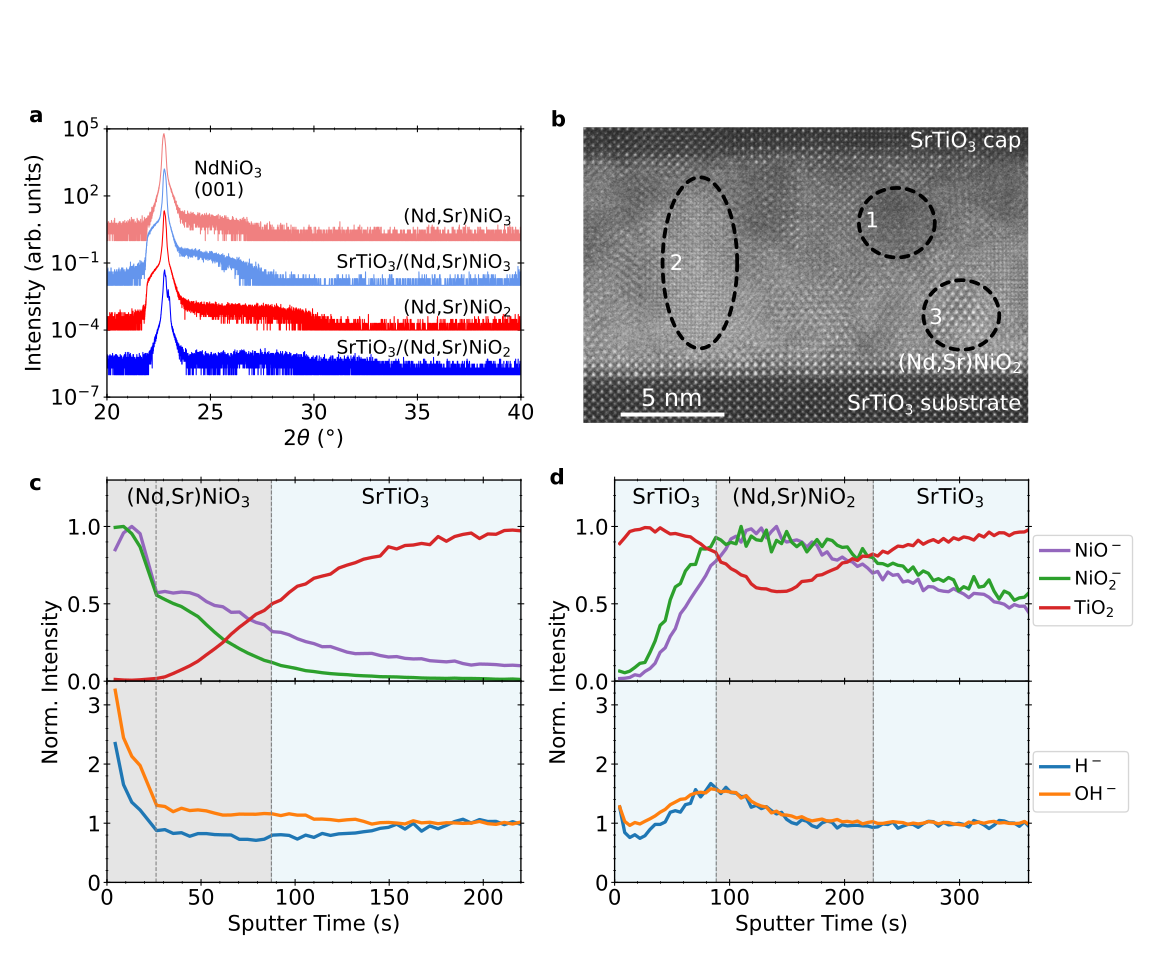}
    \caption{a. XRD on Nd$_{0.8}$Sr$_{0.2}$NiO$_3$ films grown by PLD, with and without a \qty{10}{\nano\meter} SrTiO$_3$ cap, showing a weak film (001) peak both before and after reduction, indicating a low-crystallinity as-grown film. Reduction further lowers crystallinity. b. Atomic-resolution cross-sectional HAADF-STEM micrograph from the reduced SrTiO$_3$/Nd$_{0.8}$Sr$_{0.2}$NiO$_2$ film showing amorphous (marked as 1) and crystalline regions (marked as 2 and 3). The low-crystalline quality of the film after reduction is clearly visible and agrees with the XRD data. c. SIMS depth profile of the as-grown Nd$_{0.8}$Sr$_{0.2}$NiO$_3$ film indicates a separate surface layer. d. SIMS depth profile of the reduced Nd$_{0.8}$Sr$_{0.2}$NiO$_2$ film with a SrTiO$_3$ cap, showing non-negligible but small hydrogen incorporation at the SrTiO$_3$ cap / Nd$_{0.8}$Sr$_{0.2}$NiO$_2$ interface.}
    \label{fig:capping}
\end{figure*}

We next present our findings on non-superconducting infinite-layer films Nd$_{0.8}$Sr$_{0.2}$NiO$_2$ / SrTiO$_3$ films, which are appropriately doped to result in superconductivity, but which were reduced aggressively at high temperatures (\qty{600}{\degreeCelsius} compared to $\sim$\qty{300}{\degreeCelsius}); this reduction is enough to significantly hydrogen-dope a similar perovskite material, BaZrO$_3$ \cite{orvis2019BaZrO3reduction}. Furthermore, a common practice is to cap perovskite nickelate films with SrTiO$_3$ prior to reduction to provide balanced strain for structural stability of the film throughout its entire thickness \cite{LiDiscovery2019,lee2020aspects}. We therefore compare samples with and without an SrTiO$_3$ capping layer grown \textit{in situ} on the 10 nm Nd$_{0.8}$Sr$_{0.2}$NiO$_3$ before reduction.

XRD measurements, shown in Figure \ref{fig:capping}a, reveal that the crystalline quality of both capped and uncapped films prior to reduction is lower, with broader, lower intensity film peaks. Importantly, while the (002) Nd$_{0.8}$Sr$_{0.2}$NiO$_3$ peak is sharp, the expected perovskite (001) film peak is suppressed; further higher-resolution measurements, shown in the Supplemental Fig. \ref{fig:XRDconfigs}, resolve the presence of an unidentified potential RP phase. As before, the topotactic reduction process reduces the $c$-axis lattice parameter, but the peak intensity drops dramatically. Transmission electron micrographs of these samples, such as that shown in Fig. \ref{fig:capping}b, reveal the segregation of the film into multiple crystalline phases and amorphous-like regions. Thus, the aggressive reduction of these films is non-uniform and disordered rather than controlled, resulting in increased mosaicity and a loss of crystalline quality. This is corroborated by electronic transport, as discussed further in the supplemental information, which indicates that capped and uncapped films exhibit sharply different resistivities. 

As shown in Fig. \ref{fig:capping}c, which plots SIMS measurements from the as-grown, uncapped Nd$_{0.8}$Sr$_{0.2}$NO$_3$ film, the initial transient region shows much higher yields of all ions which may indicate differences in crystallinity near the surface, potentially from the emergence of a polycrystalline scale layer in uncapped samples over time \cite{parzyck2024synthesis}. In the bulk region, the H$^-$ and OH$^-$ intensities are similar to the substrate.

Despite the significant difference in crystalline quality and reduction conditions, the effects of reduction are startlingly similar to our other observations. Figure \ref{fig:capping}d shows the integrated peak intensities in the reduced SrTiO$_3$ / Nd$_{0.8}$Sr$_{0.2}$NiO$_2$ bilayer. Once again, H$^-$ and OH$^-$ intensities are elevated at the film/cap interface -- though over a broader spatial extent -- with almost a 50\% increase over the baseline in the substrate. Thus, while the SrTiO$_3$ cap appears to trap hydrogen, this enhancement is again far below the orders of magnitude which would be expected for significant hydrogen incorporation, remaining within a factor of two of the substrate values. 

\section{Theoretical Calculations}

\begin{figure}
    \centering
    \includegraphics[width=\columnwidth]{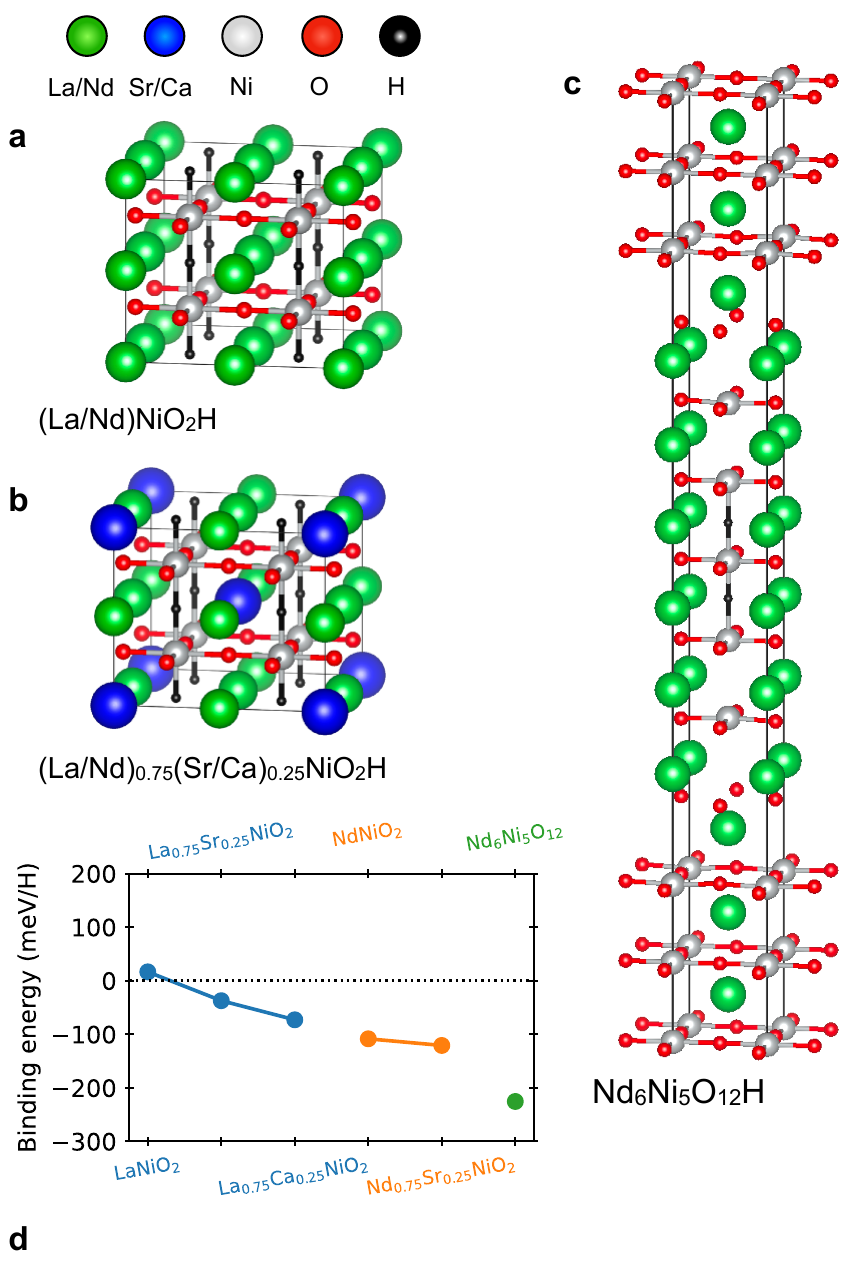}
    \caption{Topotatic-H energetics for superconducting nickelates. Crystal structures indicating the positions of topotatic-H used for (a) (La,Nd)NiO$_{2}$, (b) (La,Nd)$_{0.75}$(Sr,Ca)$_{0.25}$NiO$_{2}$, and (c) Nd$_{6}$Ni$_{5}$O$_{12}$. (d) Binding energies for superconducting layered nickalates, where a positive (negative) binding energy indicates favoring (disfavoring) H-intercalation.}
    \label{fig:topo-H-dft}
\end{figure}

To further understand the lack of hydrogen incorporation in the nickelates analyzed above via SIMS (both superconducting and non-superconducting), density-functional theory (DFT)-based calculations were performed to explore the energetics of topotactic hydrogen in infinite-layer $R$NiO$_2$ ($R$ = rare-earth, both doped and undoped) as well as in the quintuple-layer nickelate Nd$_6$Ni$_5$O$_{12}$. To investigate whether it is energetically favorable to intercalate hydrogen, we compute the hydrogen binding energy ($E_{b}$) for the topotactic process as:
\begin{equation}
%\label{eqn:eb-eqn}
E_{b} = \{ E[R\mathrm{NiO}_2] + n\times \mu[H] - E[R\mathrm{NiO}_2\mathrm{H}] \} / n,
\end{equation}
where $E[R\mathrm{NiO}_2]$ and $E[R\mathrm{NiO}_2\mathrm{H}]$ are the total energies for the infinite-layer $R$NiO$_{2}$ and hydride-oxide $R$NiO$_{2}$H compounds, $\mu[H] = E[\mathrm{H}_{2}]/2$ is the chemical potential of H, and $n$ are the number of H atoms in the (super)cell. Analogous expressions are used for $R_{0.75}$(Sr,Ca)$_{0.25}$NiO$_{2}$ and Nd$_{6}$Ni$_{5}$O$_{12}$. A positive (negative) $E_{b}$ indicates that the topotatic hydrogen intercalation is favorable (unfavorable). %and the hydride-oxide (NdNiO$_{2}$H$_{\delta}$) compound will form instead of the desired byproduct (NdNiO$_{2}$ and $H_{2}$/2). 
The calculated binding energies are summarized in Fig.~\ref{fig:topo-H-dft}. We find that the incorporation of H into $R$NiO$_{2}$, $R_{0.75}$(Sr,Ca)$_{0.25}$NiO$_{2}$, Nd$_{6}$Ni$_{5}$O$_{12}$ is systematically unfavorable, in agreement with experiments. 

%across a wide range of layered square-planar structures, which scrutinizes the possibility of hydrogen incorporation in these films by calculating the H-topotactic binding energies. Overall, we find that infinite-layer and square-planar structures, whether doped or undoped, are energetically stable against the inclusion of hydrogen. 

%Due to the difference in structure and chemistry between the all the compounds scrutinized above via SIMS, there is no \textit{a priori} reason to expect similar behavior concerning hydrogen incorporation. However, since we repeatedly find no 

In summary, we searched for hydrogen across a wide range of superconducting and non-superconducting nickelate films, with different cation and dopant chemistry, structures, growth methods, reduction conditions, and crystalline quality. Not only did we find no significant concentrations of hydrogen in superconducting films, we were also unable to use excessive reduction temperature or time to force significant amounts of hydrogen into these structures. These results are consistent with first-principles calculations which show that hydrogen incorporation is energetically unfavorable in both infinite-layer and layered square-planar compounds. At most, we observed increased concentrations by a factor of two from the trace amounts already present within the substrates. Furthermore, hydrogen, as hydride or hydroxide ions (H$^-$ and OH$^-$), was more likely to be found in SrTiO$_3$ caps or in the substrates than in the nickelate films themselves. This propensity for hydrogen to appear in higher concentrations in SrTiO$_3$ capping layers and SrTiO$_3$/nickelate interfaces is interesting in the context of recent work showing the important role such capping layers can play in facilitating the reduction process \cite{parzyck2024synthesis}.

It should be noted that our measurements generally reveal as-grown samples which are almost completely devoid of hydrogen prior to reduction. CaH$_2$ does appear to introduce hydrogen into the system, as evidenced by changes in both film and substrate levels in as-grown La$_{0.78}$Ca$_{0.22}$NiO$_3$ and reduced La$_{0.78}$Ca$_{0.22}$NiO$_2$ films, for example. However, topotactic reduction appears unable to introduce hydrogen into these nickelates at the levels near 25\% previously cited as critical doping for superconductivity \cite{DingHydrogen2023}. At such high levels of incorporated hydrogen, Ding \textit{et al.} observed H$^-$ intensities approximately 40$\times$ to 60$\times$ the substrate concentration, and approaching a factor of 200$\times$ to 600$\times$ near the film surface. We find no evidence of such large relative H$^-$ intensities in the films studied in this work.

Therefore, although superconductivity is highly sensitive to reduction optimization, this is likely due to the crystalline quality and oxygen stoichiometry, and not hydrogen stoichiometry. Of course, this study does not demonstrate that superconductivity requires the \textit{absence} of incorporated hydrogen. The results of Ding \textit{et al.} clearly demonstrate that it is possible to realize the superconducting state in samples which contain significant hydrogen concentrations. This work instead indicates that many films appear resistant to hydrogen infiltration, and most importantly, superconductivity may be readily realized without it.

%TC:ignore
\section{Experimental Methods}

\subsection{Sample Synthesis: Thin Film Deposition and Reduction}

\subsubsection{\texorpdfstring{La$_{1-x}$(Ca,Sr)$_x$NiO$_2$ Films}{La1-x(Ca,Sr)xNiO2 Films}}
\label{sec:LXNOgrowth}
Thin films, $\sim$\qty{6}{\nano\meter} thick, of the infinite-layer nickelates La$_{0.78}$Ca$_{0.22}$NiO$_2$ and La$_{0.8}$Sr$_{0.2}$NiO$_2$ were grown on SrTiO$_3$ (001) substrates using pulsed laser deposition (PLD) and CaH$_2$ topotactic reduction under conditions as previously reported \cite{chow2023pairing,chow2022paulilimit}. Notably, reduction was performed using \qty{0.1}{\gram} of CaH$_2$ powder at \qtyrange{340}{360}{\degreeCelsius} for \qty{1}{\hour} (\qty{2}{\hour} for over-reduced samples). After reduction, samples were capped \textit{ex-situ}, at room temperature, with amorphous SrTiO$_3$ to protect the surface from reoxidation.

\subsubsection{\texorpdfstring{Nd$_6$Ni$_5$O$_{12}$ and NdNiO$_2$ Films}{Nd6Ni5O12 and NdNiO2 Films}}
\label{sec:NNOgrowth}

We use ozone-assisted molecular beam epitaxy (MBE) to synthesize the precursor NdNiO$_{3}$ / SrTiO$_{3}$ (001) and Nd$_{6}$Ni$_{5}$O$_{16}$ / NdGaO$_{3}$ (110) films in Fig. \ref{fig:NNO-overreduced} and Fig. \ref{fig:SCRP}, respectively. To calibrate the nickel and neodymium elemental fluxes, we synthesize NiO on MgO (001) and Nd$_{2}$O$_{3}$ on yttria-stabilized zirconia (YSZ (111)), then measure the film thickness via x-ray reflectivity. Next, we synthesize NdNiO$_{3}$ / LaAlO$_{3}$ (001) and use the $c$-axis lattice constant and film thickness to refine the Nd / Ni ratio and monolayer dose, respectively. Using the optimized neodymium and nickel shutter times from the synthesis of NdNiO$_{3}$ / LaAlO$_{3}$, we synthesize the Ruddlesden–Popper nickelates \textit{via} monolayer shuttering. Both NdNiO$_{3}$ and Ruddlesden–Popper nickelates are synthesized at a substrate temperature of \qtyrange{500}{600}{\degreeCelsius} with $\sim1.0\times 10^{-6}$ Torr distilled ozone (Heeg Vacuum Engineering). The MBE synthesis conditions and calibration scheme are described in Refs. \cite{PanPRM2022_RPgrowth, pan2022n5RP}; similar techniques were also used in Refs. \cite{Li_YNie2020APL, Sun_YNie2021PRB}. 

The perovskite and Ruddlesden–Popper films are reduced to the square-planar phase \textit{via} CaH$_{2}$ topotactic reduction. The following methods are similar to those used elsewhere \cite{LeeAPL2020_aspects_synthesis, Li_YNieFrontiersPhys2021_cation_stoich_SC_112, pan2022n5RP}. First, the as-grown films are cut into identical pieces, and the pieces to be reduced are tightly wrapped in aluminum foil (All-Foils) to avoid direct contact between the film and CaH$_{2}$. Each film is then placed in a borosilicate tube (Chemglass Life Sciences) with $\sim$\qty{0.1}{\gram} of CaH$_{2}$ pieces ($>$92\%, Alfa Aesar). The borosilicate tube is pumped down to $<$\qty{0.5}{\milli\torr}, sealed, and then heated for several hours at $\sim$\qty{290}{\degreeCelsius} in a convection oven (Heratherm, Thermo Fisher Scientific) with a \qty{10}{\degreeCelsius\per\minute} heating rate.

\subsubsection{\texorpdfstring{Nd$_{0.8}$Sr$_{0.2}$NiO$_3$ Films}{Nd0.8Sr0.2NiO3 Films}}
\label{sec:NSNOgrowth}

\begin{figure}
    \centering
    \includegraphics[width=0.5\textwidth]{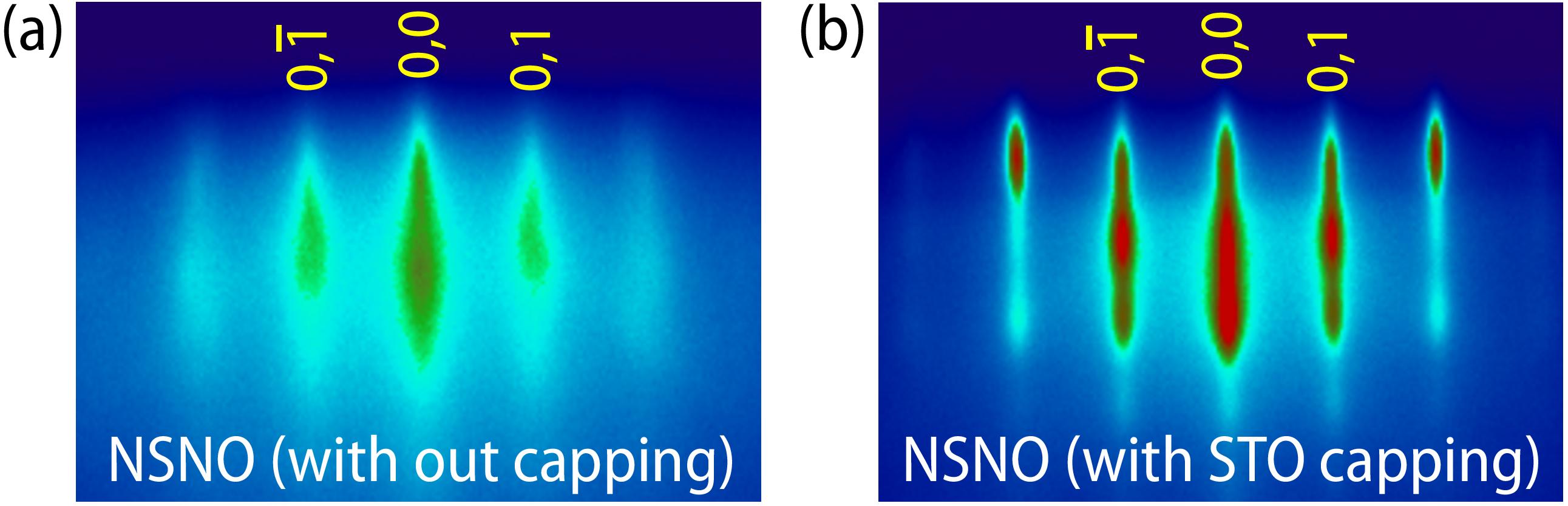}
    \caption{RHEED image of the Nd$_{0.8}$Sr$_{0.2}$NiO$_3$ (NSNO) films (a) with out and (b) with SrTiO$_3$ (STO) capping layer after cooling the samples to room temperature.}
    \label{fig:NSNORHEED}
\end{figure}

Polycrystalline targets of NdNiO$_3$ and Nd$_{0.8}$Sr$_{0.2}$NiO$_3$ were prepared by the liquid-mix technique \cite{Patel2022,patelstabilization2020}. \qty{10}{\nano\meter} thick Nd$_{0.8}$Sr$_{0.2}$NiO$_3$ films were grown on (001) SrTiO$_3$ substrates using a Neocera PLD system equipped with an \textit{in-situ} RHEED (Staib Instruments, Germany). The depositions were conducted using a KrF excimer laser operating at \qty{2}{\hertz} with a fluence of \qty{1.5}{\joule\per\square\centi\meter}. During the deposition, a dynamic oxygen pressure of \qty{150}{\milli\torr} was maintained, and the substrate temperature was \qty{735}{\degreeCelsius}. The optional \qty{10}{\nano\meter} thick SrTiO$_3$ capping layer was grown at the same condition as the film. After the deposition, all samples were \textit{in-situ} annealed at the deposition temperature in an oxygen atmosphere of \qty{500}{\torr} for \qty{30}{\minute} and subsequently cooled to room temperature at a rate of \qty{15}{\degreeCelsius\per\minute}. The streaky RHEED pattern of specular (0 0) and off-specular (0 1), (0 -1) spots, recorded after cooling the samples to room temperature confirm the excellent surface morphology of the films (Figure \ref{fig:NSNORHEED}).

The as-grown films were sealed in evacuated ($\sim$\qty{1}{\milli\torr}) quartz ampoules with \qty{0.1}{\gram} CaH$_2$ powder (90-95\%, Thermo Scientific Chemicals). The ampoules were then baked in a muffle furnace at \qty{600}{\degreeCelsius} for up to \qty{10}{\hour}. The temperature ramp rate was fixed at \qty{10}{\degreeCelsius\per\minute}. Once the ampoules were opened, the reduced films were immediately rinsed in \textit{n}-butanol and isopropanol in an ultrasonic bath for \qty{3}{\minute}.% After sonication, the films were stored in a nitrogen glove box. 

\subsection{X-ray Diffraction}

X-ray diffraction (XRD) measurements were performed at room temperature before and after reduction by each group on commercially available x-ray diffractometers using Cu K$\alpha_1$ ($\lambda$ = \qty{1.5406}{\angstrom}) radiation.

% \subsubsection{Lin Er}
% X-ray diffraction (XRD) measurements were performed at ambient conditions before and after reduction on commercially available x-ray diffractometers using a Cu K$\alpha$ wavelength and parallel-plate collimation geometry (?).

% \subsubsection{Mundy}

% X-ray diffraction (XRD) measurements were performed using a  Malvern Panalytical Empyrean diffractometer with Cu K$\alpha_1$ ($\lambda = 1.5406$ \AA) radiation. 

% \subsubsection{Ranjan}

% XRD measurements were done at room temperature by using a lab-based Rigaku Smartlab X-ray diffractometer using Cu-K$_\alpha$ radiation with a scintillator detector.

% \subsection{Reflectometry}
% \sub
\subsection{X-ray Reflectometry}

X-ray reflectometry (XRR) measurements were performed at ambient conditions in a horizontal configuration using a Rigaku SmartLab diffractometer. The incident beam was collimated using the parallel beam slit and an incident slit of \qty{30}{\micro\meter} height to improve $Q$-resolution. The Cu K$\alpha_1$ wavelength ($\lambda = $ \qty{1.5406}{\angstrom}) was isolated by using a Ge-(220)$\times$2 monochromator. The scattered beam was further collimated by a \qty{0.2}{\milli\meter} receiving slit. The data were reduced using the \emph{reductus} web-service \cite{reductus} and fit to a slab model using Refl1D \cite{kirby2012phase}.

\subsection{Time-of-Flight Secondary Ion Mass Spectroscopy}

Time-of-flight SIMS was performed using an IONTOF IV (Münster, Germany) equipped with a 20 keV Ar$^+_{2600\pm1000}$ cluster source for sputtering, a 30 keV Bi$_3^+$ liquid metal ion source for analysis, and a time-of-flight mass analyzer. Depth profiling was performed in non-interlaced mode with 1 scan of analysis with a resolution of (128 × 128) pixels, 10 scans of sputtering, and at least 0.5 s of charge compensation per cycle, where both the analysis and sputter rasters were kept inside a (500 × 500) µm area. The corresponding ion doses were 1.9 × 10\textsuperscript{9} ions/cm\textsuperscript{2} (0.12 pA) for Bi\textsubscript{3}\textsuperscript{+}, and between 2.1 × 10\textsuperscript{14} ions/cm\textsuperscript{2} to 2.6 × 10\textsuperscript{14} ions/cm\textsuperscript{2} (5.1 nA to 6.4 nA) per cycle for the cluster source due to day-to-day fluctuations in the beam current. On especially insulating samples or substrates, a small drop of silver paint was used to electrically contact the sample surface to the sample holder for further charge compensation.

For reliable detection of H$^-$ ions, contributions from residual gases were minimized by keeping the chamber pressure below \qty{5e-7}{\pascal}. Both negative and positive ions were analyzed at separate spots, and the signal rastered over multiple spots was averaged after normalizing for the highest intensity ion unique to the substrate (TiO$_2^-$ or GaO$_2^-$).

Spectra were analyzed using SurfaceLab to perform mass calibrations, identify peaks with the appropriate compounds, and extract the total integrated peak intensity as a function of sputter time. As many molecular compounds can have similar mass, peak assignments were made carefully, considering factors such as mass offset, isotopic distribution, and similarity in profile shape to other known oxide and hydroxide species.

\subsection{Electron Microscopy}

Nd$_6$Ni$_5$O$_{12}$ cross-sectional STEM specimens were prepared by the standard focussed ion beam (FIB) lift-out process on a Thermo Fisher Scientific Helios G4 UX or FEI Helios 660 FIB. HAADF-STEM was performed on a probe-corrected Thermo Fisher Scientific Spectra 300 X-CFEG operating at 300 kV with a 30 mrad convergence semi-angle. 

The reduced NiNdO$_2$ cross-sectional scanning transmission electron microscope (STEM) specimens were prepared using focused ion beam (FIB) liftout and thinning. Samples were first thinned down using an accelerating voltage of \qty{30}{\kilo\volt} with a decreasing current from \qtyrange{100}{40}{\pico\ampere}, and then with a fine polishing process using an accelerating voltage of \qty{5}{\kilo\volt} and a current of \qty{41}{\pico\ampere}. STEM-HAADF images were acquired using Thermo Fisher Scientific Titan Themis Z G3 operated at \qty{200}{\kilo\volt}. The convergence angle was \qty{18.9}{\milli\radian} and the collection angles ranged from \qtyrange{68}{280}{\milli\radian}.

The reduced SrTiO$_3$-capped Nd$_{0.8}$Sr$_{0.2}$NiO$_3$ cross-sectional STEM specimens were prepared using a Thermo Fisher Scientific Helios G4 UXe PFIB Dual Beam system, using a standard lift-out procedure. The film surface was protected by depositing protective C and W layers to prevent damage to the film surface during ion beam milling. The foil was first thinned down to $\sim$\qty{200}{\nano\meter} using the \qty{30}{\kilo\volt} ion-beam, and then with a final polishing process using a \qty{5}{\kilo\volt} ion-beam. The high-resolution STEM-HAADF imaging was carried out using a Thermo Fisher Scientific Spectra 200 operated at \qty{200}{\kilo\volt}, with a probe semi-convergence angle of \qty{25}{\milli\radian} and current of $\sim$\qty{20}{\pico\ampere}, and the collection angles ranged from \qtyrange{54}{200}{\milli\radian}. The energy dispersive X-ray spectroscopy (EDS) chemical mapping was performed using the same instrument utilizing the Bruker Dual-X X-ray detectors, using an electron beam current of $\sim$\qty{100}{\pico\ampere}.

%\section{Theoretical Methods}
\subsection{Computational methods}

Density-functional theory (DFT)-based calculations were performed to theoretically explore the energetics of topotactic hydrogen in the infinite-layer nickelate $R$NiO$_2$ ($R$= La, Nd, both doped and undoped) as well as in the quintuple-layer nickelate Nd$_6$Ni$_5$O$_{12}$.  For $R$NiO$_{2}$H$_{\delta}$, $R_{0.75}$(Sr,Ca)$_{0.25}$NiO$_{2}$H$_{\delta}$, and Nd$_{6}$Ni$_{5}$O$_{12}$H$_{\delta}$ ($\delta = 0, 1$) structural relaxations were performed using the VASP code~\cite{vasp1,vasp2,vasp3} with the the Perdew-Burke-Ernzerhof version of the generalized gradient approximation (GGA-PBE)~\cite{pbe}. For the infinite-layer materials ($R$NiO$_{2}$ and $R_{0.75}$(Sr,Ca)$_{0.25}$NiO$_{2}$) up to a $2\times2\times2$ supercell was used to accommodate the appropriate H content and/or (Sr,Ca)-doping level. GGA-PBE was chosen as it provides lattice constants in close agreement with experimental data, as shown in Supplementary Note H. A $\Gamma$-centered $13\times13\times15$ ($9\times9\times11)$ $k$-mesh was used for the $1\times1\times1$ unit cells ($2\times2\times2$ supercells) with a 0.05 eV Gaussian smearing. For Nd$_{6}$Ni$_{5}$O$_{12}$, a $\Gamma$-centered $9\times9\times9$ $k$-mesh with a 0.05 eV Gaussian smearing was used. The size of the plane-wave basis sets was set with a kinetic energy cut-off of 520 eV. For $R=\mathrm{Nd}$, we have used a pseudopotential where the Nd($4f$) electrons are frozen in the core. To compute the chemical potential of hydrogen ($\mu[\mathrm{H}]$), we optimized an H$_{2}$ dimer in $15^{3}$ \AA{}$^{3}$ box with energy cutoff set to 325 eV.

%The binding energy ($E_{b}$) for the topotatic hydrogen process is computed as~\cite{SiChainsTheory2023, SiTopotactic2020}:
%\begin{equation}
%\label{eqn:eb-eqn}
%E_{b} = \{ E[\mathrm{NdNiO}_2] + n\times \mu[H] - %E[\mathrm{NdNiO}_2\mathrm{H}_{\delta}] \} / n,
%\end{equation}
%where $E[\mathrm{NdNiO}_2]$ and $E[\mathrm{NdNiO}_2\mathrm{H}_{\delta}]$ are the total energies for the infinite-layer NdNiO$_{2}$ and hydride-oxide NdNiO$_{2}$H$_{\delta}$ compounds, $\mu[H] = E[\mathrm{H}_{2}]/2$ is the chemical potential of H, and $n$ are the number of H atoms in the (super)cell. Analogous expressions are used for Nd$_{0.75}$Sr$_{0.25}$NiO$_{2}$ and Nd$_{6}$Ni$_{5}$O$_{12}$. A positive (negative) $E_{b}$ indicates that the topotatic hydrogen process is favorable (unfavorable) and the hydride-oxide (NdNiO$_{2}$H$_{\delta}$) compound will form instead of the desired byproduct (NdNiO$_{2}$ and $H_{2}$/2). To compute the chemical potential of hydrogen ($\mu[\mathrm{H}]$), we optimize a H$_{2}$ dimer in $15^{3}$ \AA{}$^{3}$ box with energy cutoff set to 325 eV.

\section*{Author Contributions}

PPB, S. Muramoto, and AJG performed and analyzed ToF-SIMS measurements. PPB, AJG, and PQ performed XRR and analyzed and interpreted all reflectometry data. LEC deposited and reduced La$_{0.8}$Sr$_{0.2}$NiO$_3$/La$_{0.8}$Sr$_{0.2}$NiO$_2$ and La$_{0.78}$Ca$_{0.22}$NiO$_3$/La$_{0.78}$Ca$_{0.22}$NiO$_2$ films, and performed electronic characterization. NdNiO$_3$ and Nd$_6$Ni$_5$O$_{12}$ films were fabricated deposited by GAP, DFS, and QS, while GAP and DFS performed reduction and electronic characterization. TEM of Nd$_6$Ni$_5$O$_{12}$ and NdNiO$_2$ films were performed by BHG and YTS, respectively. Nd$_{0.8}$Sr$_{0.2}$NiO$_3$ films were deposited and initially characterized by RKP and S. Middey, reduced by MS, and imaged \textit{via} TEM by KJ and YTS. HL and ASB performed theoretical calculations for all film compositions. The study was designed by AJG, PPB, PQ, DFS, JAM, AA, and JR. The paper was written by PPB and AJG with input from all authors.

\begin{acknowledgments}
We thank Kyuho Lee for insightful discussions. We also thank Kerry Sieben for x-ray assistance. We thank Hanjong Paik for supporting the growth of the $n$ = 5 superconducting sample.

Certain commercial equipment, instruments, software, or materials are identified in this paper in order to specify the experimental procedure adequately. Such identifications are not intended to imply recommendation or endorsement by NIST, nor it is intended to imply that the materials or equipment identified are necessarily the best available for the purpose. 

Research was performed in part at the NIST Center for Nanoscale Science and Technology. Electron microscopy was performed at the Cornell Center for Materials Research (CCMR) Shared Facilities, which are supported by the NSF MRSEC Program (No. DMR-1719875). B.H.G. and L.F.K acknowledge support from PARADIM and the Packard Foundation. S. Middey acknowledges MHRD, Government of India for financial support under the STARS research funding scheme (grant number: STARS/APR2019/PS/156/FS). PPB and PQ received funding from the NRC RAP. JAM, ASB, and HL acknowledge support from NSF grant no. DMR-2323971. JR and MS acknowledge support from an ARO MURI program with award no. W911NF-21-1-0327, and the National Science Foundation of the United States under grant number DMR-2122071. Materials growth and electron microscopy were supported in part by the Platform for the Accelerated Realization, Analysis, and Discovery of Interface Materials (PARADIM) under NSF Cooperative Agreement no. DMR-2039380. KJ and YTS acknowledge support from USC Viterbi startup funding and the USC Research and Innovation Instrumentation Award. Electron microscopy data were acquired at the Core Center of Excellence in Nano Imaging at USC. LEC and AA acknowledge support from the Ministry of Education (MOE), Singapore, under its Tier-2 Academic Research Fund (AcRF), Grants No. MOET2EP50121-0018 and MOE-T2EP50123-0013.
\end{acknowledgments}

% \newpage

% \appendix{Supplemental Information}
\section{Supplemental Information}
\renewcommand{\thefigure}{S\arabic{figure}}
\setcounter{figure}{0}

\subsection{Sample Summary}

Table \ref{tab:sampletable} lists details of all samples referenced in this paper, along with their fabrication details and which figures they correspond to.

\begin{table*}
\centering
\begin{tabular*}{1\textwidth}{|c | r | c | l|}
    \hline
    Label & Material & Synthesis & Figures\\
    \hline\hline
    As-grown (La,Ca)NiO$_3$ & 5.4 nm La$_{0.78}$Ca$_{0.22}$NiO$_3$ / SrTiO$_3$ & \ref{sec:LXNOgrowth} & \ref{fig:SCperovskite}b,c,d, \ref{fig:LCNOSIMSall}\\
    Superconducting (La,Ca)NiO$_2$ & 29 nm SrTiO$_3$ / 4 nm La$_{0.78}$Ca$_{0.22}$NiO$_2$ / SrTiO$_3$ & & \ref{fig:SCperovskite}a,b,c,d,e,f, \ref{fig:LCNOSIMSall}\\
    Over-reduced (La,Ca)NiO$_2$ & 3.2 nm SrTiO$_3$ / 3.5 nm La$_{0.78}$Ca$_{0.22}$NiO$_2$ / SrTiO$_3$ & & \ref{fig:LCNOSIMSall}\\
    \hline
    As-grown (La,Sr)NiO$_3$ & 5.3 nm La$_{0.8}$Sr$_{0.2}$NiO$_3$ / SrTiO$_3$ & & \ref{fig:SCperovskite}b, \ref{fig:LSNOSIMSall}\\
    Superconducting (La,Sr)NiO$_2$ & 6 nm SrTiO$_3$ / 4.2 nm La$_{0.8}$Sr$_{0.2}$NiO$_2$ / SrTiO$_3$ & & \ref{fig:SCperovskite}a,b,g,h, \ref{fig:SIMSisotopes}, \ref{fig:LSNOSIMSall}\\
    Over-reduced (La,Sr)NiO$_2$ & 31 nm SrTiO$_3$ / 4.4 nm La$_{0.8}$Sr$_{0.2}$NiO$_2$ / SrTiO$_3$ & &\\
    \hline\hline
    Superconducting Nd$_6$Ni$_5$O$_{12}$ & 100 nm Pt / 10 nm Ti / 23 nm Nd$_6$Ni$_5$O$_{12}$ / NdGaO$_3$ & \ref{sec:NNOgrowth} & \ref{fig:schematics}c, \ref{fig:SCRP}a,b,c,d, \ref{fig:RPSIMSall}\\
    \hline\hline
    As-grown NdNiO$_3$ & 17.3 nm NdNiO$_3$ / SrTiO$_3$ & \ref{sec:NNOgrowth} & \ref{fig:NNO-overreduced}a,c,d, \ref{fig:NNOall}, \ref{fig:NNOtransport}\\
    Reduced NdNiO$_2$ & 17.3 nm NdNiO$_2$ / SrTiO$_3$ & & \ref{fig:NNO-overreduced}a,e,f, \ref{fig:SIMScarbon}, \ref{fig:NNOall}, \ref{fig:NNOtransport}\\
    \hline
    Reduced NdNiO$_2$ & 16 nm NdNiO$_2$ / SrTiO$_3$ & & \ref{fig:NNO-overreduced}b\\
    \hline\hline
    As-grown (Nd,Sr)NiO$_3$ & 10 nm Nd$_{0.8}$Sr$_{0.2}$NiO$_3$ / SrTiO$_3$ & \ref{sec:NSNOgrowth} & \ref{fig:capping}a,c,d, \ref{fig:NSNORHEED}a, \ref{fig:NSNOall}, \ref{fig:XRDconfigs}\\
    \hline
    As-grown capped (Nd,Sr)NiO$_3$ & 10 nm SrTiO$_3$ / 10 nm Nd$_{0.8}$Sr$_{0.2}$NiO$_3$ / SrTiO$_3$ & & \ref{fig:capping}a, \ref{fig:NSNORHEED}b, \ref{fig:XRDconfigs}\\
    \hline
    Reduced (Nd,Sr)NiO$_2$ & 10 nm Nd$_{0.8}$Sr$_{0.2}$NiO$_2$ / SrTiO$_3$ & & \ref{fig:capping}a, \ref{fig:XRDconfigs}\\
    \hline
    Reduced capped (Nd,Sr)NiO$_2$ & 10 nm SrTiO$_3$ / 10 nm Nd$_{0.8}$Sr$_{0.2}$NiO$_2$ / SrTiO$_3$ & & \ref{fig:capping}b,e,f, \ref{fig:NSNOall}, \ref{fig:XRDconfigs}, \ref{fig:NSNOTEM}, \ref{fig:NSNOEELS}\\
    \hline
    \hline
    Mica & (KF)$_2$(Al$_2$O$_3$)$_3$(SiO$_2$)$_6$(H$_2$O) & MTI Corp.& \ref{fig:mica}\\
    \hline
\end{tabular*}
\label{tab:sampletable}
\caption{Table of samples and where they are referenced in this paper. Grouped samples are pieces split from the same deposited film.}
\end{table*}

\subsection{Hydrogen Quantification using SIMS}
\label{sec:quantification}

Although ToF-SIMS is very sensitive to mass and depth, extracting quantitative information (e.g., chemical stoichiometries) can be challenging due to so-called matrix effects where the chemical and electronic environment of the material (matrix) affects the ionization of the sputtered species. First, the choice of the sputtering beam can change the matrix. For example, the use of a Cs$^+$ sputter beam results in Cs implantation into the film, lowering the work function and dramatically increasing the yield of negative ions \cite{wehbe2014xps}. Because different ions can interact differently with the implanted Cs, the ion yield ratios can vary drastically. To minimize potential chemical interactions with the sample, we chose to use an Ar$^+$ noble gas cluster source for sputtering, which is also volatile and escapes to vacuum after bombardment. The type of analysis beam (Bi$_3$$^+$, O$_2$$^+$, Ga$^+$, etc.) also has an effect on ionization, and this needs to be specified when comparing quantitative results. 

\begin{figure*}
    \centering
    \includegraphics[width=\textwidth]{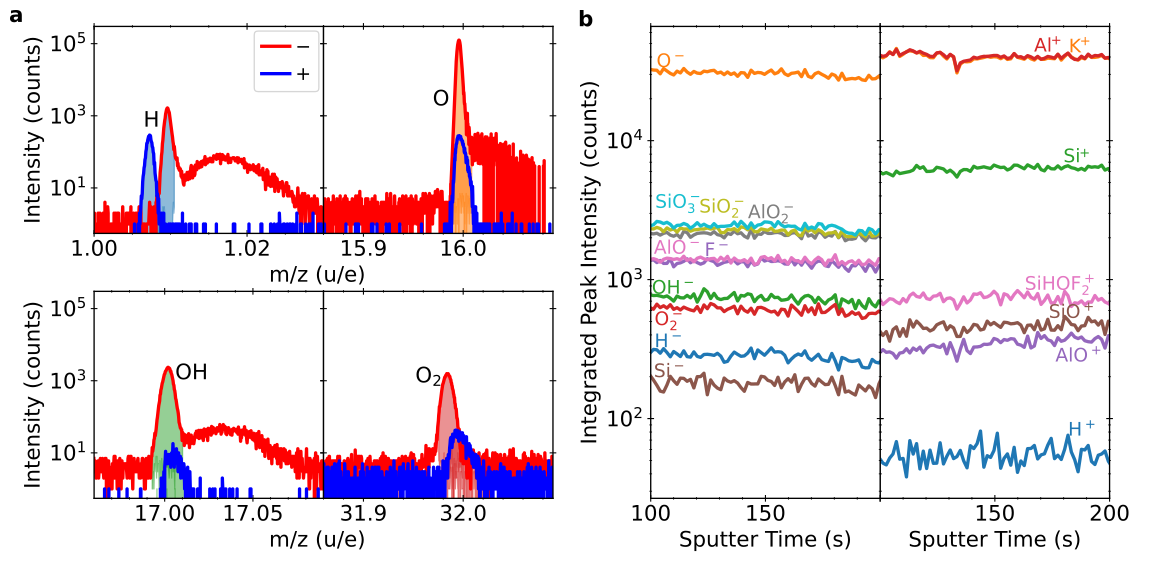}
    \caption{a. SIMS showing the spectral peak shapes of  H$^\pm$, O$^\pm$, OH$^\pm$, and O$_2^\pm$ ions. The broad peaks centered around $m/z=1.02$ and $m/z=17.03$ are attributed to charging effects, and were not included for the construction of the depth profile. b. SIMS depth profiles showing the integrated peak intensities as a function of sputtered depth for both elemental and molecular ions from a mica substrate.}
    \label{fig:mica}
\end{figure*}

Second, quantification requires matrix-matched standards. Since even small impurities can affect ionization, standards need to be prepared with known ratios of chemical species \emph{and} exist in a similar chemical environment to that of the films of interest \cite{stevie2008quantification}. Relative sensitivity factors can then be calculated to scale the integrated peak areas to atomic concentrations. We first attempt to quantify hydrogen content in our nickelate films by comparing to a mica standard, as used in Ding \textit{et al.} \cite{DingHydrogen2023}. The chemical formula for the mica (muscovite) crystals we used is (KF)$_2$(Al$_2$O$_3$)$_3$(SiO$_2$)$_6$(H$_2$O). Peak shapes as well as intensity measurements are shown in Fig. \ref{fig:mica}.

To illustrate the importance of matrix-matched standards, we compare the levels of oxygen (O$^-$ and O$_2^-$) measured in the mica and in the SrTiO$_3$ and NdGaO$_3$ substrates prior to any film reduction. Although they have comparable oxygen number densities, we found that dissimilarities in the ionization efficiencies produced drastically different results (i.e., estimated stoichiometric ratios). A quick calculation revealed that the apparent oxygen concentration was significantly lower in the mica than in the SrTiO$_3$ and NdGaO$_3$ substrates, and this is due to the presence of different elements in the matrix. The same effect is observed for all elements of the periodic table, including H, which highlights the difficulty of quantification using SIMS. In addition, hydrogen is present in adventitious carbon contamination at surfaces and interfaces, and requires careful interpretation of the profile data.

Another complication is the effect of the reduction process on the chemical and physical properties of the film, which directly affects the sensitivity factor and the sputtering rate, respectively. A lower sputter rate means that fewer number of elements and molecules are being ejected at any given time, and even though the ionization efficiency may not have changed, the intensity appears to be lower. In our study, this apparent intensity of various molecular species within a single nickelate sample set varied dramatically as a function of reduction time and conditions, as discussed in further detail in Section \ref{sec:allSIMS}. This renders even samples cleaved from a single growth and reduced differently to not be quantitatively comparable.

\subsection{Disentangling Extrinsic and Intrinsic Hydrogen}
\label{sec:adsorbates}

\begin{figure}
    \centering
    \includegraphics[width=\columnwidth]{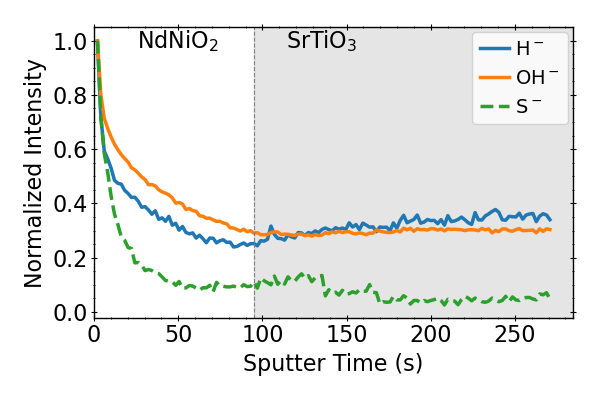}
    \caption{Comparison of ToF-SIMS hydrogen signals in reduced non-superconducting NdNiO$_2$ with surface adsorbate signals. Intensities have been normalized to the surface value. The slower decay of hydrogen intensities indicate that some of the hydrogen signal is intrinsic to the sample.}
    \label{fig:SIMScarbon}
\end{figure}
As mentioned earlier, SIMS can also detect surface adsorbates, and requires a way to distinguish hydrogen intrinsic to the sample from surface contamination. Surface adsorbates are often hydrocarbons, so we can compare carbon peaks such as C$_2^-$, as well as other contaminant peaks, and see if hydrogen trends with these compounds. The rapid decay in signal at the surface is usually due to these hydrocarbons, and their contribution to the H and OH signals can be subtracted from those inside the film.
\begin{figure}
    \centering
    \includegraphics[width=\columnwidth]{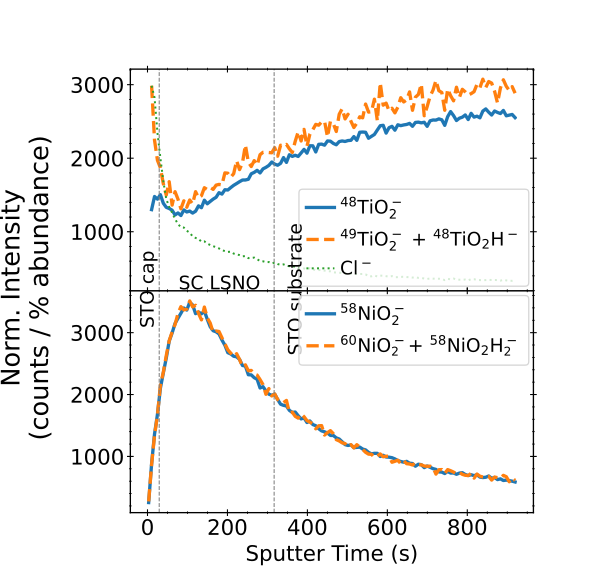}
    \caption{Comparison of the isotopic distributions of Ti and Ni oxides in superconducting La$_{0.8}$Sr$_{0.2}$NiO$_2$. The less abundant isotopic oxides have an overlapping mass with the more abundant isotopic hydroxides. By scaling the depth profiles by the natural isotopic abundance, we can separate out these contributions. In this case, there is a contribution from the Ti hydroxide, but no evidence of Ni hydroxide. The sharp decay in peak intensity for $^{48}$TiO$_2$H$^-$ at the surface indicates that the Ti oxide of the SrTiO$_3$ cap has adsorbed some hydrogen, but that has not affected the nickelate film.}
    \label{fig:SIMSisotopes}
\end{figure}
We can further confidently attribute certain signals to measured hydrogen by isotope analysis. Metal oxides such as Ti will have the same ionization efficiency regardless of the isotope (e.g. $^{48}$TiO$_2$ vs. $^{49}$TiO$_2$). Species with isobaric interference, or those that overlap in mass, such as $^{49}$TiO$_2$ and $^{48}$TiO$_2$H, can be distinguished by analyzing the natural isotopic distribution in the mass spectrum or profile shape, as shown in Fig. \ref{fig:SIMSisotopes}. In this figure, the less abundant $^{49}$TiO$_2$$^-$ has an overlapping mass with $^{48}$TiO$_2$H$^-$ at $m/z$ = 64.945 that cannot be resolved. However, by constructing the depth profile and seeing their profile shapes, it is possible to identify which molecule the peak corresponds to. In this case, there is a contribution from the Ti hydroxide. Similarly for Ni oxides, comparison of the profiles show that $^{60}$NiO$_2$$^-$ is the likely assignment for this peak instead of the hydroxide. The sharp decay in peak intensity for $^{48}$TiO$_2$H$^-$ at the surface indicates that the Ti oxide of the SrTiO$_3$ cap has adsorbed some hydrogen, but that has not affected the nickelate film.

\subsection{Previous detection of hydrogen using SIMS}

\begin{figure}
    \centering
    \includegraphics[width=0.3\textwidth]{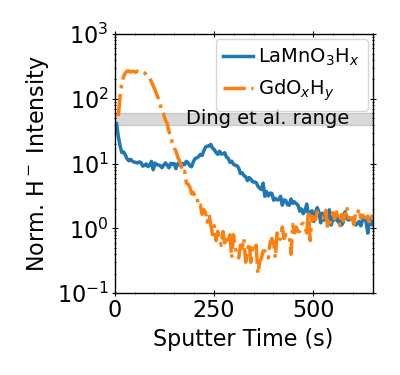}
    \caption{Examples of previously measured and published SIMS depth profiles for the H$^-$ ion in the hydrogen containing oxides LaMnO$_3$H$_x$ and GdO$_x$H$_y$ \cite{lu2022engineering,sheffels2023insight}, normalized to 1 in the substrate.}
    \label{fig:OldSIMSBenchmark}
\end{figure}

In order to place our observations into context, it is helpful to consider previous studies which used the ToF-SIMS instrument employed in this study to examine oxide films with large degrees of hydrogen incorporation, such as humidity-treated GdO$_x$H$_y$, LaMnO$_3$ membranes exposed to water, and CaH$_2$-reduced BaZrO$_3$\cite{sheffels2023insight, lu2022engineering, orvis2019BaZrO3reduction}. In Fig. \ref{fig:OldSIMSBenchmark}, we show reproduced H$^-$ intensity curves from a LaMnO$_3$ freestanding membrane with incorporated water due to the dissolution of Sr$_3$Al$_2$O$_6$ during processing\cite{lu2022engineering}. Also shown is a film of GdO$_x$H$_y$ which is very near the Gd(OH)$_3$ composition\cite{sheffels2023insight}. In these systems, we found enhancements of the H$^+$, H$^-$, and OH$^{-}$ intensities which were 1 to 3 orders of magnitude above the substrate background, and were able to establish the high hydrogen concentrations based on the effects on the material properties. This prior experience provides reasonable upper and lower boundaries for expected effects and demonstrates our sensitivity to intrinsic hydrogen. Interestingly, the results from Ding \textit{et al.} fell between these limits. We conclude therefore that hydrogen would readily have been detected in our measurements of layered square-planar nickelate samples if it were present at previously reported levels.

% \subsection{Attempt to quantify hydrogen using a reference material}

\subsection{Extended SIMS Characterization}
\label{sec:allSIMS}

\begin{figure*}
    \centering
    \includegraphics[width=\textwidth]{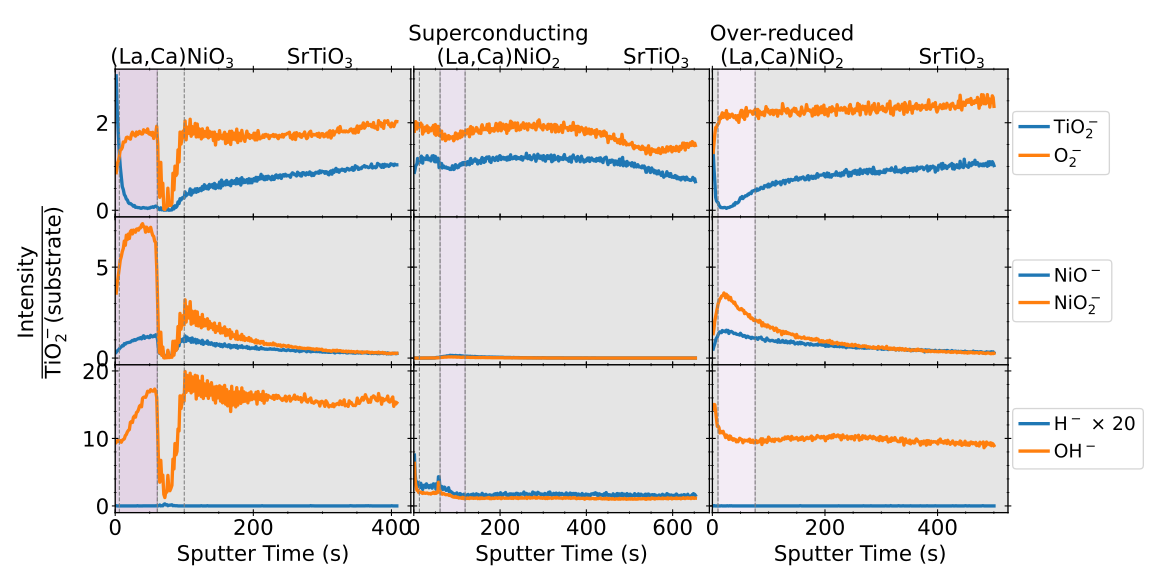}
    \caption{ToF-SIMS depth profiles for the as-grown, superconducting and over-reduced La$_{0.78}$Ca$_{0.22}$NiO$_{3\rightarrow2}$  films, including those shown in \ref{fig:SCperovskite}.}
    \label{fig:LCNOSIMSall}
\end{figure*}

\begin{figure*}
    \centering
    \includegraphics[width=\textwidth]{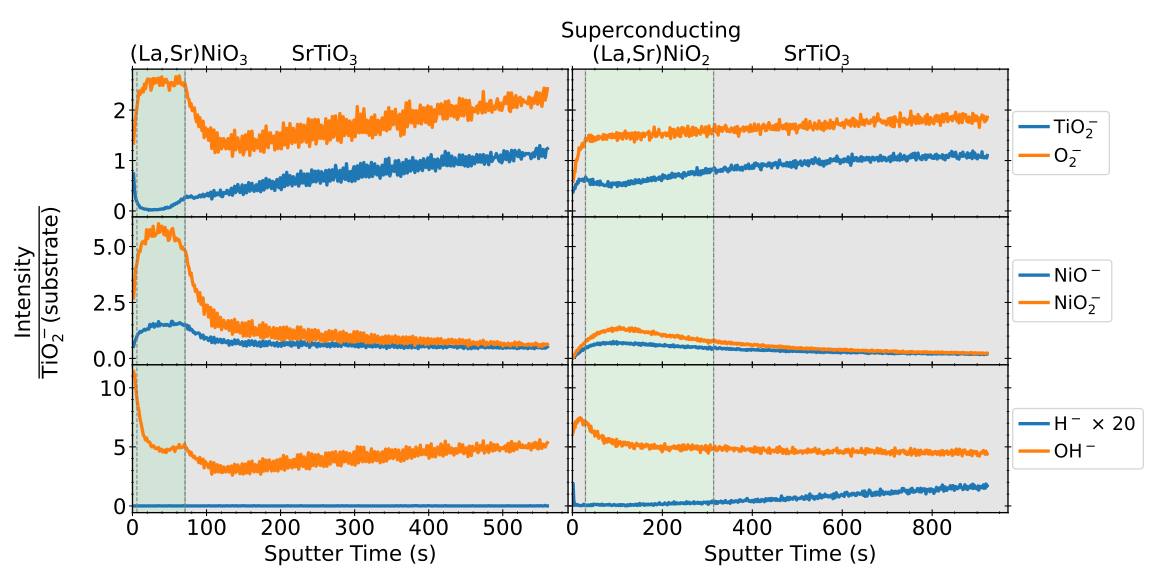}
    \caption{ToF-SIMS depth profiles for the as-grown and superconducting La$_{0.8}$Sr$_{0.2}$NiO$_{3\rightarrow2}$ films including, those shown in \ref{fig:SCperovskite}.}
    \label{fig:LSNOSIMSall}
\end{figure*}

\begin{figure*}
    \centering
    \includegraphics[width=\textwidth]{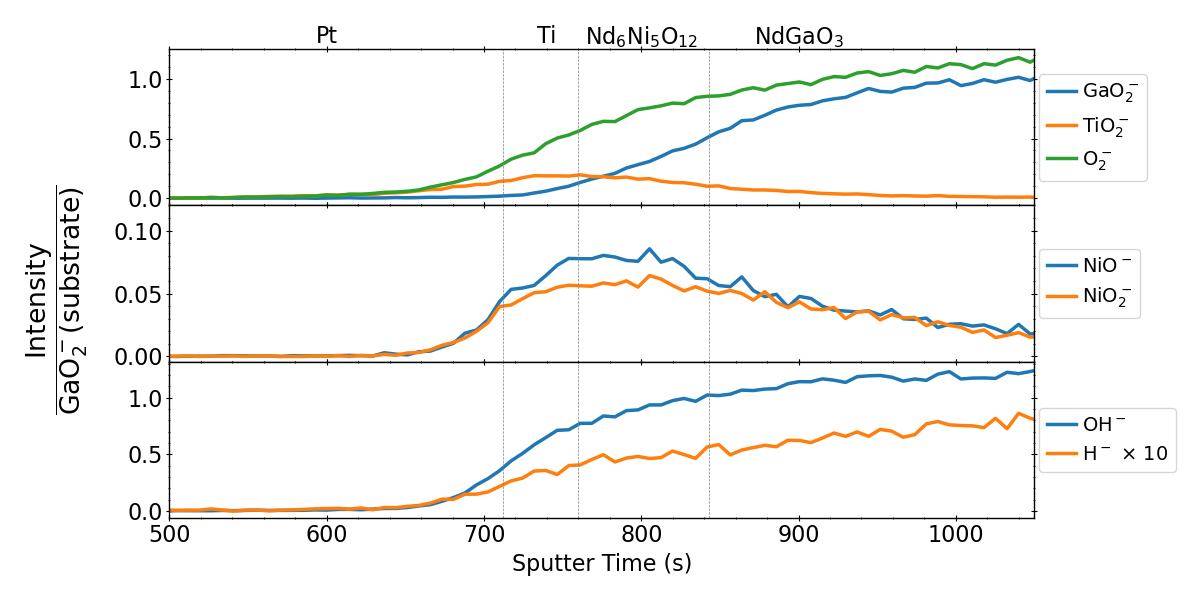}
    \caption{ToF-SIMS depth profile for the Nd$_6$Ni$_5$O$_{12}$ film shown in \ref{fig:SCRP}. Intensities are normalized to the steady-state GaO$_2^-$ in the NdGaO$_3$ substrate.}
    \label{fig:RPSIMSall}
\end{figure*}

\begin{figure*}
    \centering
    \includegraphics[width=\textwidth]{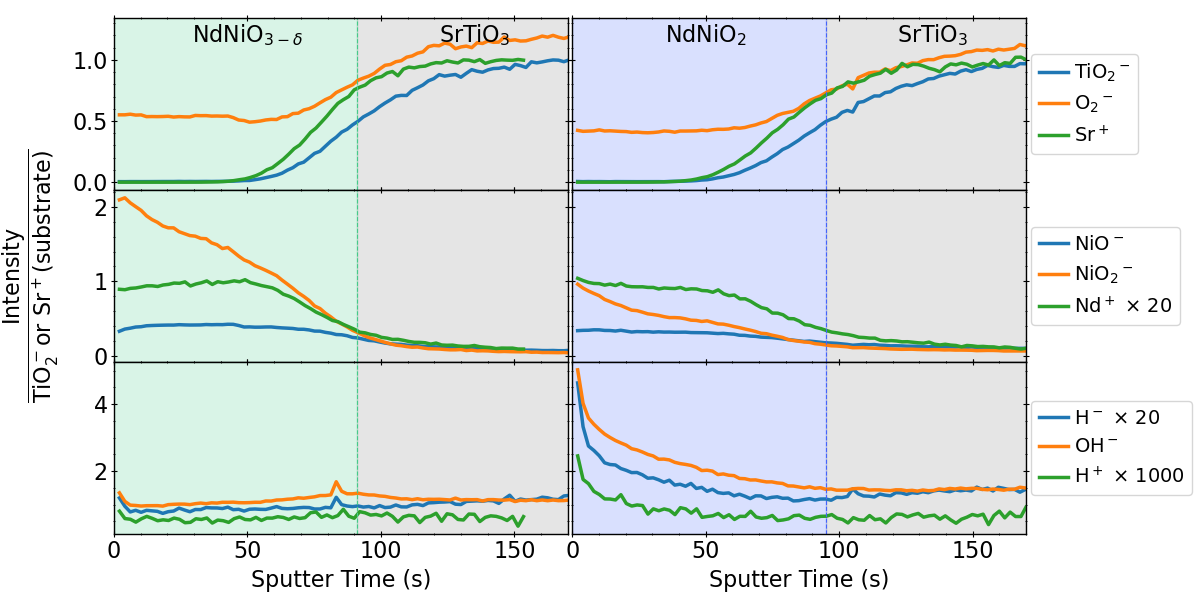}
    \caption{ToF-SIMS depth profiles for the NdNiO$_3$ and non-superconducting NdNiO$_2$ films shown in \ref{fig:NNO-overreduced}. Intensities are normalized to steady-state TiO$_2^-$ or Sr$^+$ in the SrTiO$_3$ substrate.}
    \label{fig:NNOall}
\end{figure*}

\begin{figure*}
    \centering
    \includegraphics[width=\textwidth]{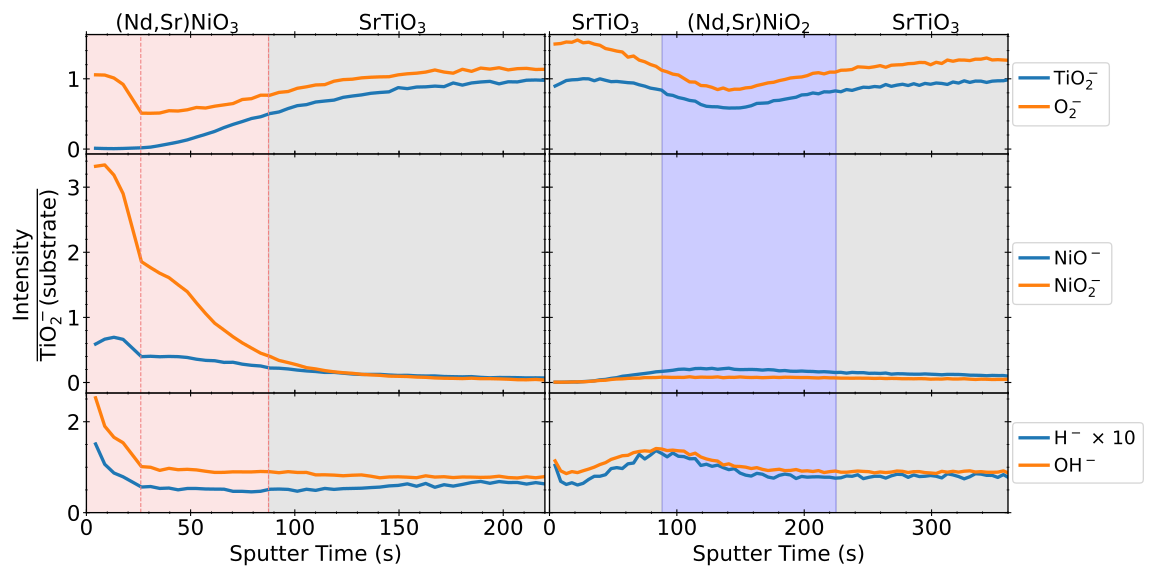}
    \caption{ToF-SIMS depth profiles for the Nd$_{0.8}$Sr$_{0.2}$NiO$_3$ and SrTiO$_3$/Nd$_{0.8}$Sr$_{0.2}$NiO$_2$ shown in \ref{fig:capping}. Intensities are normalized to steady-state TiO$_2^-$ in the substrate.}
    \label{fig:NSNOall}
\end{figure*}

For each sample, we have replotted SIMS depth profiles by scaling all intensities to an appropriate steady-state ion in the substrate to compare between sample types. Certain compounds with much higher (OH$^-$) or lower (H$^-$) intensities have been scaled for visualization clarity. These are shown in Figures \ref{fig:LCNOSIMSall}, \ref{fig:LSNOSIMSall}, \ref{fig:RPSIMSall}, \ref{fig:NNOall}, and \ref{fig:NSNOall}.

The layer interfaces are identified as halfway up the transitions of ions unique to each layer (TiO$_2^-$ in the SrTiO$_3$, NiO$^-$ in the nickelate film, GaO$_2^-$ in the NdGaO$_3$), noting that heavier ions have broader depth profiles due to redeposition into the sputter crater. We also note that the sputter rate is faster for lower crystallinity layers, including amorphous SrTiO$_3$ caps, as well as lower-crystallinity reduced films. In fact, attempted measurements of the aggressively reduced Nd$_{0.8}$Sr$_{0.2}$NiO$_2$ without protection of the SrTiO$_3$ cap sputtered through the film so quickly that no depth profile could be obtained.

SIMS reveals dramatic differences between as-grown, superconducting, and non-superconducting reduced samples.

First, changes in O$_2^-$ intensity within the films can be seen after reduction, confirming preferential removal of oxygen from the nickelate layers using CaH$_2$.

Next, the NiO$^-$ and NiO$_2^-$ intensities and their ratio vary dramatically between samples, even when originating from the same as-grown sample. In particular, though the intensities of these nickel oxides drops after reduction, they are not monotonic with reduction. In particular, we see that the TiO$_2$-normalized intensities are an order of magnitude lower in the superconducting La$_{0.78}$Ca$_{0.22}$NiO$_2$ but increase again in the over-reduced film (Fig. \ref{fig:LCNOSIMSall}). Furthermore, this effect is not seen in all samples (Fig. \ref{fig:NNOall}). We conclude that the Ni valence and electronic state significantly affect the measured intensities of many ions \cite{SIMSEuOxidation,SIMSCeOxidation}, making even quantitative sample-to-sample comparisons difficult.

Finally, though we see no evidence for extensive hydrogen incorporation in any of these samples, the hydrogen signals which are measured are primarily present as OH$^-$, and secondarily H$^-$. The H$^+$ peak is orders of magnitude smaller, and in many cases undetectable.

\subsection{Electrical characterization of non-superconducting films}

\begin{figure}
    \centering
    \includegraphics[width=0.5\textwidth]{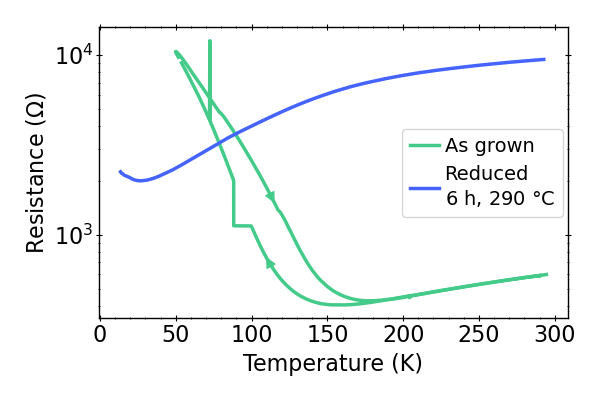}
    \caption{Two-point resistance measurements of as-grown and reduced NdNiO$_{3\rightarrow 2}$ shown in Fig. \ref{fig:NNO-overreduced}.}
    \label{fig:NNOtransport}
\end{figure}

Fig. \ref{fig:NNOtransport} shows resistance vs. temperature for the non-superconducting, undoped NdNiO$_3$/SrTiO$_3$ shown in Fig. \ref{fig:NNO-overreduced}. These electrical transport measurements were taken using a home-built electrical ``dipstick probe" compatible with a helium dewar. Indium contacts were soldered on the four corners of each sample in a van der Pauw (VDP) configuration. AC transport measurements were taken at 17.777 Hz using an SR-830 lock-in amplifier. The voltage and current were measured simultaneously to determine the resistance. Note that values shown are the measured resistance of a single VDP channel of these \qty{5}{\milli\meter}$\times$\qty{10}{\milli\meter} samples, not sheet resistance or resistivity. %, and so absolute scaling is irrelevant, but temperature trends can still be evaluated.

The original as-grown film shows the hysteretic metal-insulator transition expected for a stoichiometric NdNiO$_3$ film, while the reduced infinite-layer film exhibits the metallic behavior expected for undoped NdNiO$_2$.

Nd$_{0.8}$Sr$_{0.2}$NiO$_2$ films were insulating enough that two-point resistances were measured only at room temperature. As-grown, without significant growth optimization, uncapped films started with resistances between \qtyrange{500}{700}{\kilo\ohm}, while capped films started with resistances between \qtyrange{6}{8}{\kilo\ohm}. After reduction, the resistance of uncapped films actually increased and was not measurable using a standard multimeter, while capped films had a lower resistance.

\subsection{Further structural characterization}

\begin{figure*}
    \centering
    \includegraphics[width=.8\textwidth]{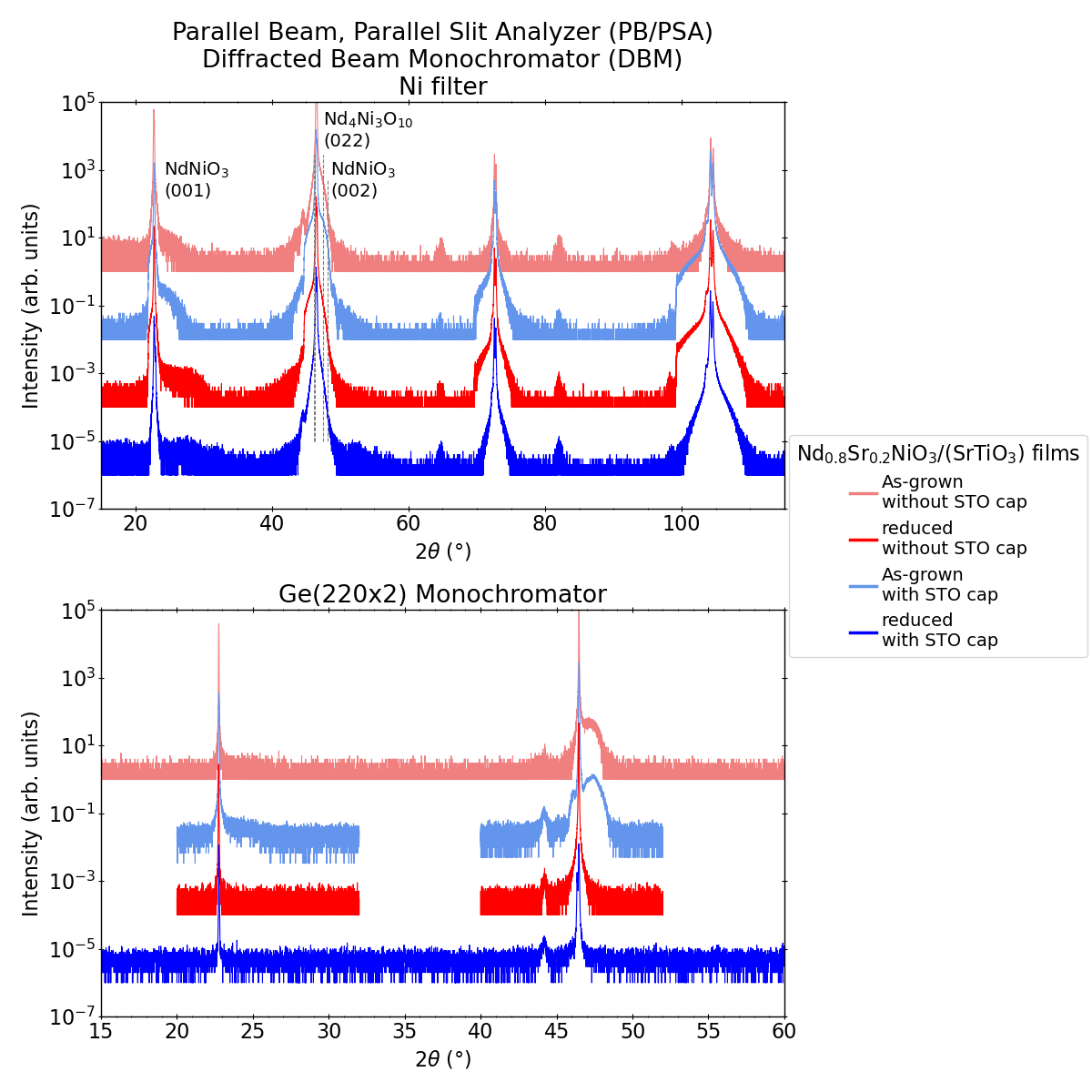}
    \caption{Full XRD scans of samples in Figure \ref{fig:capping}a taken using two instrument configurations. On top are high-intensity, low-resolution measurements used to survey for impurity phases, while on bottom are the same scans taken using a lower-intensity, high-resolution mode.% \textcolor{red}{Need to add as-grown into here}
    }
    \label{fig:XRDconfigs}
\end{figure*}

Figure \ref{fig:XRDconfigs} shows complementary XRD measurements of the capped and uncapped Nd$_{0.8}$Sr$_{0.2}$NiO$_{3\rightarrow2}$ films taken using two configurations of a Rigaku SmartLab XRD. The top scans were measured in a high-intensity mode, using a parallel beam (PB) selection slit, and a parallel slit analyzer (PSA). Cu K$\alpha$ lines were selected using a Ni filter. The lighter color scans were measured in a high-resolution mode, using a Ge (220)$\times$2 monochromator to select Cu K$\alpha_1$ radiation on the incident side. The Diffracted Beam Monochromator assembly was installed on the receiving side containing a Ni filter, Ge (400)$\times$2 analyzer, and \qty{0.5}{\degree} PSA. While the high-resolution measurements allow us to measure the as-grown NSNO (002) film peak at \qty{47.485}{\degree}, and confirms the much lower intensity of the corresponding (001) peak, the peaks of the reduced film are so low in intensity that they cannot be seen at all.

\begin{figure*}
    \centering
    \includegraphics[width=\textwidth]{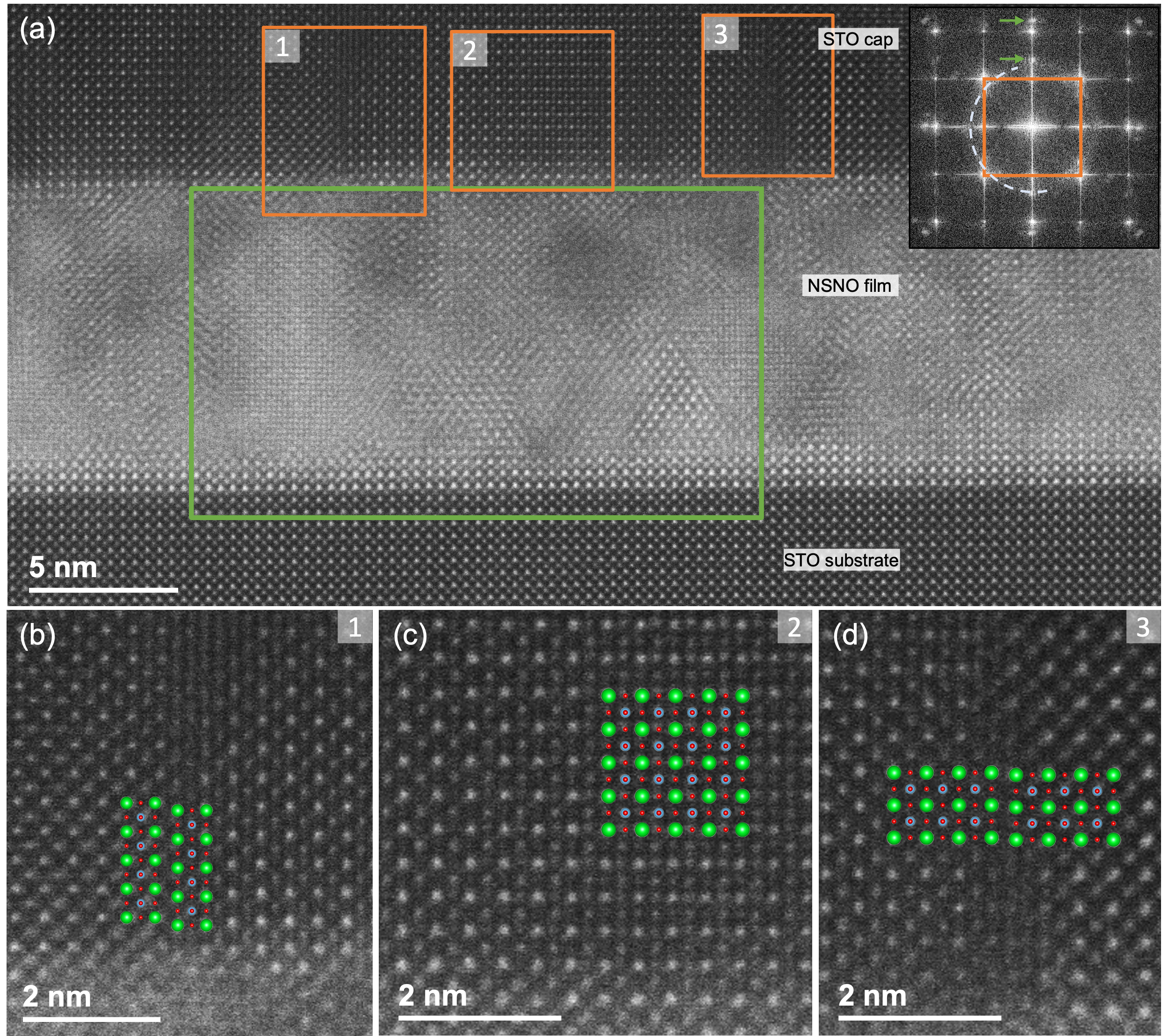}
    \caption{(a) The larger field of view of the image shown in \ref{fig:capping}b, showing the presence of defects in the STO capping layer as well. The inset shows the FFT pattern of the image. The spots corresponding to the cubic STO substrate (and cap) is denoted by the square, the extra spots and the diffused ring indicate the presence of additional crystalline and amorphous regions in the films, as respectively shown by the arrows and dashed semi-circle. (b-d) Magnified regions of the SrTiO$_3$ capping layer showing the presence of various forms of defects, with a schematic structural model overlaid (Sr = green, Ti = blue, O = red spheres). (b) Region 1 shows a Ruddlesden-Popper-like phase at the nickelate-cap interface. (c) Region 2 shows strong signal from the oxygen columns, which cannot be only due to O as low Z elements cannot to scatter to the high collection angle used in this experiment. (d) Region 3 shows the planes in the SrTiO$_3$ cap are highly distorted.
    }
    \label{fig:NSNOTEM}
\end{figure*}

\begin{figure*}
    \centering
    \includegraphics[width=\textwidth]{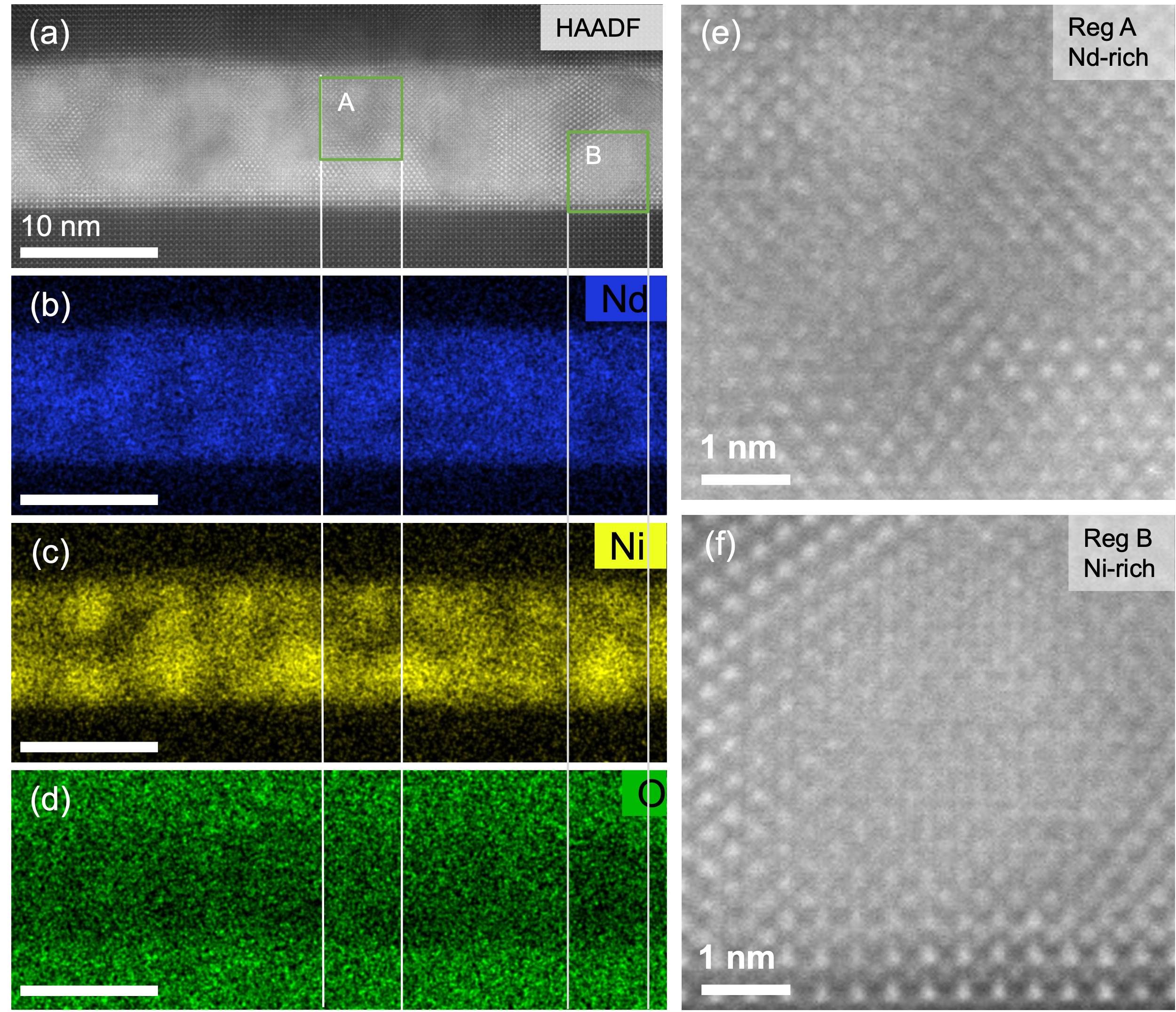}
    \caption{(a-d) HAADF-STEM image and the corresponding EDS elemental maps obtained from the reduced SrTiO$_3$-capped (Nd,Sr)NiO$_2$ film. A clear segregation of Ni and Nd is observed in the nickelate film. (e, f) The amorphous-like and crystalline domains correspond to the Nd-rich and Ni-rich region, respectively.
    }
    \label{fig:NSNOEELS}
\end{figure*}

The crystalline quality and nature of defects in the reduced SrTiO$_3$-capped Nd$_{0.8}$Sr$_{0.2}$NiO$_2$ films is studied by cross-sectional HAADF-STEM studies and is observed to be highly defective as shown in Fig. \ref{fig:capping}b. Mosaicity is observed in the (Nd,Sr)NiO$_2$ layer, appearing as a mixture of amorphous-like and crystalline domains, marked by the circles in \ref{fig:capping}b. Particularly, two different atomic arrangements deviating from the projected [100] perovskite structure is observed in the crystalline regions marked as 2 and 3 in \ref{fig:capping}b. Similar defects are present uniformly throughout the observed thin region ($\sim$\qty{5}{\micro\meter}) of the FIB foil. The Fast Fourier Transform (FFT) of the atomic resolution HAADF-STEM image is shown in the inset of Fig. \ref{fig:NSNOTEM}a. The sharp spots with the 4-fold symmetry corresponds to the SrTiO$_3$ perovskite structure along the [100] orientation. The FFT also shows the presence of additional spots corresponding to periodicities deviating from the STO structure (marked by the green arrows in inset of Fig. \ref{fig:NSNOTEM}a), indicating the occurrence of a phase transformation, most likely during the reduction process. The presence of weak diffused rings corresponds to the presence of the amorphous regions (marked by the dashed circle in inset of Fig. \ref{fig:NSNOTEM}a). The SrTiO$_3$ capping layer is also observed to consist of defects as shown in Fig. \ref{fig:NSNOTEM}b-d.

Further, the chemical nature of the film was studied using STEM-EDS mapping. The elemental maps show chemical segregation in the reduced SrTiO$_3$/Nd$_{0.8}$Sr$_{0.2}$NiO$_2$ film in the form of Ni-rich and Nd-rich regions and shown in Fig. \ref{fig:NSNOEELS}a-d. By correlating the chemical maps with the high-resolution HAADF image, it is observed that the Nd-rich regions correspond to the amorphous-like regions, while the Ni-rich region correspond to the crystalline regions. Finally, Sr was not observed to be concentrated in this film. The STEM results thus provide a direct proof for the crystalline and chemical disorder in the Nd$_{0.8}$Sr$_{0.2}$NiO$_2$ film after the aggressive reduction process.

\subsection{\label{sec:dft}Additional DFT data}
\begin{figure}
    \centering
    \includegraphics[width=\columnwidth]{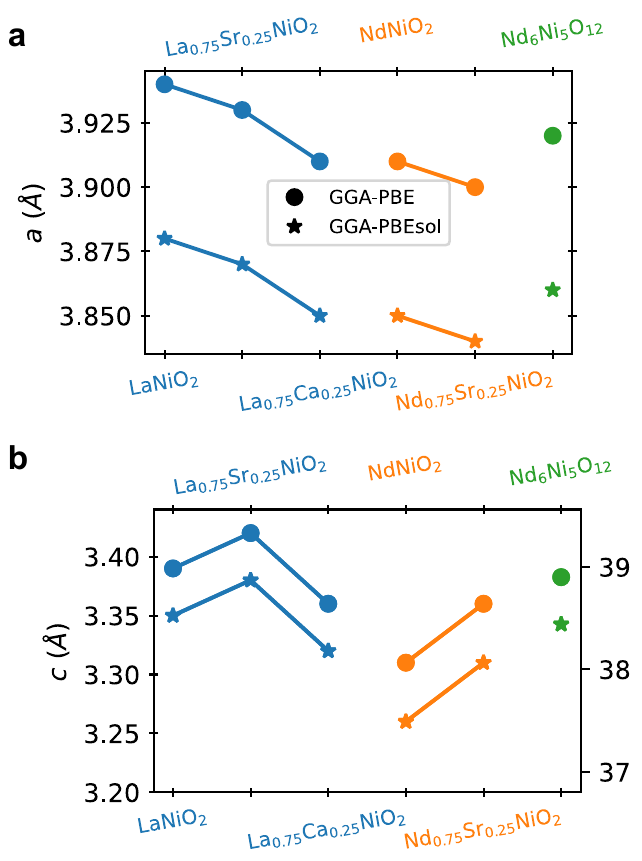}
    \caption{Structural data from the DFT-optimized structures. (a) In-plane ($a$) lattice constants. (b) Out-of-plane ($c$) lattice constants (right axis corresponds to Nd$_{6}$Ni${5}$O$_{12}$).}
    \label{fig:dft-lattice}
\end{figure}

We provide here the structural data obtained from DFT-based structural relaxations for two choices of exchange-correlation functional: GGA-PBE and GGA-PBEsol. The optimized lattice constants for all the nickelates considered in the main text within these two functionals are summarized in Fig.~\ref{fig:dft-lattice}. We find good agreement with the experimental lattice constants for the undoped infinite-layer compounds ($R$ = La, Nd) is obtained with the GGA-PBE functional, which serves as a benchmark for our DFT-optimized lattice constants ($a_{exp}$= 3.92 \AA, $c_{exp}$= 3.28 \AA ~for NdNiO$_2$ and $a_{exp}$=3.96 \AA, $c_{exp}$= 3.38 \AA ~for LaNiO$_2$ \cite{Hayward1999, HAYWARD_nd}). Previous works have used PBEsol to obtain H-binding energies in infinite-layer nickelates~\cite{SiChainsTheory2023, SiTopotactic2020}, but we find that the GGA-PBEsol functional gives rise to optimized lattice constants that are quite far from the experimental values. 

\clearpage
%\section*{References}
\bibliography{references}

%TC:endignore

% \bibliography{apssamp}% Produces the bibliography via BibTeX.

\end{document}